\documentclass[final, 5p, sort&compress, times, twocolumn]{elsarticle}

\usepackage{framed} 
\usepackage{multicol} 
\usepackage{nomencl} 
\usepackage{afterpage}
\makenomenclature
\renewcommand*\nompreamble{\begin{multicols}{2}}
\renewcommand*\nompostamble{\end{multicols}}
\usepackage{etoolbox}
\renewcommand\nomgroup[1]{%
 \item[\itshape
 \ifstrequal{#1}{A}{Sets}{%
 \ifstrequal{#1}{B}{Variables}{%
 \ifstrequal{#1}{C}{Parameters}{%
 \ifstrequal{#1}{E}{Abbreviations}{
 }}}}%
]}

\usepackage[numbers]{natbib}
\usepackage{tabularx}

\usepackage{makecell}
\usepackage{cellspace}
\usepackage[justification=justified,font=footnotesize,skip=0pt]{caption}
\usepackage[binary-units=true]{siunitx}
\DeclareSIUnit \VAr {VAr} 
\DeclareSIUnit \VA {VA} 
\DeclareSIUnit \rad {Radians} 

\usepackage{mathtools}
\usepackage{algorithm, algorithmic}
\usepackage[font=footnotesize]{caption,subfig}
\usepackage{multirow}

\usepackage{enumitem}
\usepackage{xcolor}
\usepackage{lipsum}
\usepackage{comment}
\allowdisplaybreaks

\usepackage{stackengine,scalerel}

\newcommand\obullet[1]{\ThisStyle{\ensurestackMath{%
  \stackon[1pt]{\SavedStyle#1}{\SavedStyle\kern.6\LMpt\bullet}}}}
\newcommand\ocirc[1]{\ThisStyle{\ensurestackMath{%
  \stackon[1pt]{\SavedStyle#1}{\SavedStyle\kern.6\LMpt\circ}}}}

\usepackage{cleveref}
\crefname{equation}{equation}{equations}
\Crefname{equation}{Equation}{Equations}
\crefrangelabelformat{equation}{(#3#1#4--#5#2#6)}

\crefmultiformat{equation}{equations (#2#1#3}{, #2#1#3)}{#2#1#3}{#2#1#3}
\Crefmultiformat{equation}{Equations (#2#1#3}{, #2#1#3)}{#2#1#3}{#2#1#3}

\usepackage{amsmath}
\usepackage{tikz}
\usepackage{mathdots}
\usepackage{yhmath}
\usepackage{cancel}
\usepackage{color}

\usepackage{array}
\usepackage{amssymb}
\usepackage{gensymb}
\usepackage{tabularx}
\usepackage{booktabs}
\usetikzlibrary{fadings}
\usetikzlibrary{patterns}
\usetikzlibrary{shadows.blur}

\def\tsc#1{\csdef{#1}{\textsc{\lowercase{#1}}\xspace}}
\tsc{WGM}
\tsc{QE}
\tsc{EP}
\tsc{PMS}
\tsc{BEC}
\tsc{DE}

\begin{document}

\begin{frontmatter}


\title{Fair Coordination {\color{black}of Distributed Energy Resources} with Volt-Var Control and PV Curtailment}                      

\author[1]{Daniel Gebbran\corref{cor1}}
\address[1]{School of Electrical and Information Engineering, The University of Sydney, Sydney, Australia}
\cortext[cor1]{Corresponding author}
\ead{daniel.gebbran@sydney.edu.au}

\author[2]{Sleiman Mhanna}
\ead{sleiman.mhanna@unimelb.edu.au}

\address[2]{Department of Electrical and Electronic Engineering, The University of Melbourne, Melbourne, Australia}

\author[1]{Yiju Ma}
\ead{yiju.ma@sydney.edu.au}

\author[3]{Archie C. Chapman}
\ead{archie.chapman@uq.edu.au}

\author[1]{Gregor Verbi\v{c}}
\ead{gregor.verbic@sydney.edu.au}

\address[3]{School of Information Technology and Electrical Engineering, The University of Queensland, Brisbane, Australia}

\begin{abstract}
This paper presents a novel distributed optimal power flow (DOPF) method for fair distributed energy resource (DER) coordination in the context of mandated rooftop PV inverter control modes.
In practice, inverter reactive power control is increasingly required by grid connection codes, which often unfairly curtail PV generation of prosumers towards the end of low-voltage feeders. 
Similarly, optimization-based DER coordination methods that aim solely for technically-efficient DER coordination do not consider the distribution of PV curtailment across customers. 
To address these concerns, we develop a tractable multi-objective DOPF method for optimal DER coordination that (i) curtails PV generation fairly across prosumers, and (ii) incorporates a standard piecewise-linear volt-var control reactive power control function without using integer variables. 
Three equity principles representing different interpretations of fairness are implemented in our coordination method; namely, egalitarian, proportional and uniform dynamic PV curtailment redistribution. 
The performance of our approach is demonstrated on low-voltage distribution feeders of different sizes (5, 10, 25, 50 and 100 prosumers) using two network topologies: line topology without lateral spurs and tree topology with lateral spurs. Each network considers three levels of PV penetration, giving 30 test systems in total.
The results demonstrate the effectiveness of the proposed DOPF method for fair DER coordination: PV curtailment is equitably distributed among prosumers with a computational burden on par with conventional DOPF approaches. Moreover, different fairness methods result in different patterns of curtailment, which a regulator may choose between.
\end{abstract}


\end{frontmatter}

\begin{table*}[!t]  
\begin{framed}
\printnomenclature
\end{framed}
\end{table*}

\nomenclature[A,8]{$\mathcal{X}$}{Feasible set of network variables for OPF.}
\nomenclature[A,a1]{$\mathcal{Z}_h$}{Feasible set of variables for prosumer $h$ for OPF.}
\nomenclature[A,9]{$\hat{\mathcal{X}}$}{Feasible set of network variables for DOPF.}
\nomenclature[A,a2]{$\hat{\mathcal{Z}}_h$}{Feasible set of variables for prosumer $h$ for DOPF.}
\nomenclature[A,6]{$\mathcal{H}$}{Set of prosumers in the power network.}
\nomenclature[A,7]{$\mathcal{T}$}{Time horizon of the problem.}
\nomenclature[A,5]{$\mathcal{B}$}{Set of buses in the power network.}

\nomenclature[A,1]{$\boldsymbol{x}$}{Set of network variables for OPF.}
\nomenclature[A,3]{$\boldsymbol{z}_h$}{Set of prosumer variables for OPF.}
\nomenclature[A,2]{$\hat{\boldsymbol{x}}$}{Set of network variables for DOPF.}
\nomenclature[A,4]{$\hat{\boldsymbol{z}}_h$}{Set of prosumer variables for DOPF.}

\nomenclature[B,1]{${q}_{\text{g},0,t}$}{Active power ($\SI{}{\kilo\watt}$) from reference bus at time $t$.}
\nomenclature[B,2]{$q_{\text{g},0,t}$}{Reactive power ($\SI{}{\kilo\VAr}$) from reference bus at $t$.}
\nomenclature[B,3]{$v_{i,t}$}{Voltage (p.u.) at bus $i$ at $t$.}
\nomenclature[B,4]{$\theta_{ij,t}$}{Voltage phase angle between buses $i$ and $j$ at $t$.}
\nomenclature[B,5]{$p_{h,t}/q_{h,t}$}{Active/reactive power demand ($\SI{}{\kilo\watt}/\SI{}{\kilo\VAr}$) of prosumer $h$ at $t$.}
\nomenclature[B,6]{$v_{h,t}$}{Voltage (p.u.) at prosumer $h$ at $t$.}
\nomenclature[B,7]{$p^{\text{bat}}_{h,t}$}{Battery active power ($\SI{}{\kilo\watt}$) of prosumer $h$ at $t$.}
\nomenclature[B,8]{$p^{\text{PV}}_{h,t}$}{PV active power generation ($\SI{}{\kilo\watt}$) of prosumer $h$ at $t$.}
\nomenclature[B,9]{$y_{h,t}$}{PV curtailment ($\SI{}{\kilo\watt}$) of prosumer $h$ at $t$.}
\nomenclature[B,a1]{$p_{h,t}^{\text{ch}}/p_{h,t}^{\text{ch}}$}{Charging/discharging of prosumer's $h$ battery at $t$.}
\nomenclature[B,a2]{$SoC_{h,t}$}{State-of-charge ($\SI{}{\kilo\watt\hour}$) of prosumer's $h$ battery, at $t$.}
\nomenclature[B,a3]{$\overline{y}_{t}$}{Maximum amount of PVC ($\SI{}{\kilo\watt}$) across all prosumers, at $t$.}
\nomenclature[B,a4]{$\underline{y}_{t}$}{Minimum amount of exported power ($\SI{}{\kilo\watt}$) across all participating prosumers, at $t$.}
\nomenclature[B,b5]{$\lambda^{\bullet}_{h,t}$}{Dual variable of prosumer $h$ at $t$, associated with given $\bullet$ (voltage, active or reactive power).}

\nomenclature[C,1]{$b_{ij}$}{Susceptance (p.u.) of branch $ij$.}
\nomenclature[C,2]{$g_{ij}$}{Conductance (p.u.) of branch $ij$.}
\nomenclature[C,3]{$c_{0}/c_{1}/c_{2}$}{Constant ($\SI{}{\$}$), linear ($\SI{}{\$\per\kilo\watt}$) and quadratic ($\SI{}{\$\per\kilo\watt\squared}$) coefficient terms of the network cost function.}
\nomenclature[C,6]{$v_{\text{r},t}/\theta_{\text{r},t}$}{Reference voltage (magnitude and angle), in p.u.}
\nomenclature[C,7]{$c^{\text{ToU}}_{t}$}{Time-of-use tariff ($\SI{}{\$\per\kilo\watt}$) at time interval $t$ of prosumer cost functions.}
\nomenclature[C,8]{$c^{\text{FiT}}$}{Feed-in-tariff ($\SI{}{\$\per\kilo\watt}$) of prosumer cost functions.}
\nomenclature[C,9]{$\overline{s}_{h}$}{Apparent power rating ($\SI{}{\kilo\VA}$) of prosumer's $h$ inverter.}
\nomenclature[C,a1]{$\overline{q}_{h}$}{Reactive power set rating ($\SI{}{\kilo\VAr}$) of prosumer's $h$ inverter.}
\nomenclature[C,a2]{${V}_{\bullet}$}{Voltage setting ($\SI{}{\volt}$) of Volt-Var Control equation.}
\nomenclature[C,a3]{$p_{h,t}^{\text{d}}$}{Active power demand ($\SI{}{\kilo\watt}$) of prosumer $h$ at $t$.}
\nomenclature[C,a4]{$\tilde{p}_{h,t}^{\text{PV}}$}{PV power generation ($\SI{}{\kilo\watt}$) of prosumer $h$ at $t$.}
\nomenclature[C,a5]{$\eta_{h}^{\bullet}$}{Battery charging/discharging efficiency of prosumer $h$.}
\nomenclature[C,a6]{$\Delta t$}{Time interval ($\SI{}{\hour}$) within $\mathcal{T}$.}
\nomenclature[C,a7]{$k$}{Iteration number.}
\nomenclature[C,a8]{$\rho_{\bullet}$}{ADMM penalty parameter.}
\nomenclature[C,a9]{$\alpha$}{Weighting coefficient for PVC penalty function.}
\nomenclature[C,b1]{$\beta$}{Weighting coefficient for reactive power VVC relaxation.}
\nomenclature[C,b2]{$\gamma$}{Positive small number to avoid division by 0.}
\nomenclature[C,b3]{$\epsilon^{\text{abs}}/\epsilon^{\text{rel}}$}{Absolute and relative feasibility tolerances for ADMM termination criteria.}
\nomenclature[C,b4]{$\mu/\tau$}{ADMM residual balancing parameters.}
\nomenclature[C,b6]{$\underline{\bullet}/\overline{\bullet}$}{Minimum/maximum magnitude operator.}
\nomenclature[C,b7]{$\bullet^{+}/\bullet^{-}$}{Non-negative terms of given $\bullet = \bullet^{+} - \bullet^{-}$.}
\nomenclature[C,b8]{${\lVert \bullet \rVert}_2$}{2-norm (Euclidean distance) of a set.}
\nomenclature[C,b9]{$\bullet^{k}$}{Previous variables values, used in current ADMM iteration $k$.}
\nomenclature[C,c1]{$\bullet^{k+1}$}{Calculated variables values obtained in current ADMM iteration $k$.}

\nomenclature[E]{AC}{Alternating current.}
\nomenclature[E]{ADMM}{Alternating direction method of multipliers.}
\nomenclature[E]{DER}{Distributed energy resources.}
\nomenclature[E]{DNSP}{Distribution network service provider.}
\nomenclature[E]{DOPF}{Distributed optimal power flow.}
\nomenclature[E]{KKT}{\emph{Karush-Kuhn-Tucker}.}
\nomenclature[E]{MINLP}{Mixed-integer nonlinear programming.}
\nomenclature[E]{OPF}{Optimal power flow.}
\nomenclature[E]{PV}{Photovoltaic.}
\nomenclature[E]{PVC}{Photovoltaic curtailment.}
\nomenclature[E]{PWL}{Piecewise-linear.}
\nomenclature[E]{SOC}{Second-order cone.}
\nomenclature[E]{VVC}{Volt-Var Control.}


\section{Introduction} \label{Sec:Introduction}
%


The centralized power system operation paradigm is rapidly shifting towards a more distributed one, driven by an increasing uptake of prosumer-owned distributed energy resources (DER), which necessitates new tools for DER management \cite{OpenEnergyNetworks}. In this context, distributed AC optimal power flow (DOPF) approaches \cite{Scott2019, Andrianesis2019, Attarha2020, Gebbran_SGES} have emerged as an effective tool for DER coordination, for three main reasons: (i) they explicitly consider network constraints, (ii) they permit a prosumer-based decomposition, which preserves prosumer prerogative and privacy, and (iii) they are computationally scalable. However, the current DOPF approaches fail to consider reactive power compensation, which is increasingly required by grid connection codes, and tend to curtail more PV generation from prosumers at the end of low-voltage feeders.

\subsection{Background} \label{Sec:Intro_Background}

The AC optimal power flow problem is typically solved using interior-point methods (IPM). Although they only guarantee local optimality, IPM are desirable as they can converge to \emph{feasible} solutions. In contrast, convex relaxations often converge to infeasible solutions \cite{Molzahn2017}, but are also desirable as they provide a lower bound on the optimal solution, which can be used to assess the quality of the solution obtained from IPM. However, the AC OPF problem quickly becomes intractable when considering a large number of DER and the longer optimization horizons required to accommodate inter-temporal couplings (e.g. battery storage). This therefore encouraged distributed solution approaches. 

{\color{black} Conceptually, distributed approaches decompose the optimization problem into several smaller subproblems, where each subproblem can be solved by an individual agent. The solution is reached by iteratively exchanging messages between the agents until the problem converges. Distributed optimization approaches can be hierarchical or fully distributed. The former include a higher level central coordinator, whereas the latter have no central entity.} 
{\color{black}Distributed optimization techniques can be broadly categorized into: (i) methods based on Lagrangian decomposition, and (ii) techniques for solving the Karush-Kuhn-Tucker conditions in a distributed manner. We provide a brief overview of each category next, but interested readers are referred to \cite{Molzahn2017} for a comprehensive survey of distributed optimization and control algorithms applied to power systems, including mathematical formulations and convergence properties.}

\subsubsection{\color{black}{Lagrangian-based methods}}
{\color{black}Lagrangian-based methods use the Lagrangian function of an optimization problem with a separable structure. 
A decomposition is obtained by duplicating (consensus) variables shared between subproblems, e.g., voltages and/or power flows. 
Consistency between the copies is enforced by additional equality constraints.}

{\color{black} The simplest Lagrangian-based method is \textit{dual decomposition} \cite{Everett_1963}. The basic idea is to \textit{relax} the coupling constraints by multiplying them by Lagrange multipliers (dual variables) and adding them to the objective. The solution procedure then iterates between minimization of the Lagrangian over the primal variables and maximization over the dual variables in a procedure called \textit{dual ascent}. Because of the decomposable structure, the minimization of the Lagrangian can be parallelized; however the dual update needs to be computed centrally. Convergence of dual ascent is often poor because the dual can be non-smooth.
	
The poor convergence of the dual decomposition can be improved by \emph{augmenting} the Lagrangian with a quadratic penalty on the consensus constraints, which smooths the Lagrangian but destroys the separability of the problem. The most popular Augmented Lagrangian-based method is the \emph{alternating direction method of multipliers} (ADMM) \cite{Boyd2011}, which uses alternate minimizations over the duplicated variables, which restores the separability.
}

{\color{black}More recently, the \emph{Augmented Lagrangian alternating direction inexact Newton} (ALADIN) method was proposed to solve distributed optimization problems \cite{Engelmann2019}. Similar to ADMM, ALADIN combines a coordination step between agents after solving local optimization problems based on the Augmented Lagrangian. The coordination step, however, entails solving an equality-constrained quadratic program, obtained from sensitivity evaluations of all the subproblems (gradients of local objective functions, Hessian approximations of local objective functions and constraints, and Jacobians of constraints). In part due to this improved coordination step, ALADIN offers locally quadratic convergence properties and may converge faster than ADMM, even for non-convex OPF \cite{Engelmann2019}.}

{\color{black}In a different approach, the \emph{analytical target cascading} (ATC) method \cite{Tosserams_2006} makes use of a hierarchical separation in a tree structure, where parent and child subproblems share coupling variables. The coupling constraints are enforced by a quadratic penalty relaxed function. Furthermore, it is possible to use the Augmented Lagrangian modeled within the ATC to be solved at each subproblem, effectively solving the ATC hierarchical subproblems using ADMM.}

{\color{black}The \emph{auxiliary problem principle} (APP) \cite{Cohen_1980} technique also decomposes an optimization problem into smaller parts, solving the subproblems based on the Augmented Lagrangian and sharing the common variables, and eventually ensuring consensus between them. However, the APP uses linearized cross-terms when modeling the Augmented Lagrangian, which simplifies it and } 
{\color{black}decouples the subproblems in a way that does not require a central coordinator.}

\subsubsection{{\color{black}KKT conditions-based methods}}

{\color{black}An alternative approach to Lagrangian relaxation is the distributed solution of the Karush-Kuhn-Tucker (KKT) optimality conditions.
In the \emph{optimality condition decomposition} (OCD) \cite{Conejo_2002}, each agent performs one Newton step to the KKT conditions for its subproblem in each iteration and shares the resulting primal and dual variables with the neighboring agents. Each primal and dual variable is assigned to a specific subproblem, where all other variables are fixed at their last update. The coupling for these fixed variables is modeled using linear penalties, added to the objective function, which does not require a central coordinator \cite{Molzahn2017}.}

{\color{black}The \emph{consensus+innovation} (C+I) algorithm \cite{Kar_2014} also solves the KKT conditions in a distributed manner without a central coordinator, but allows all variables in each subproblem to vary. Each update takes into account the coupling between the neighboring Lagrange multipliers, in tandem with a local \emph{innovation} term, enforcing the demand and supply power balance. The limit point of the algorithm satisfies the KKT conditions.}

\subsection{{\color{black}Decomposition of the Optimal Power Flow}}
Different decomposition approaches to solve the distributed OPF problem have been studied for more than 20 years \cite{Kim1997}. 
They range from region-wise decompositions \cite{Erseghe2014,Guo_2017} to more recent component-wise decomposition \cite{Peng2018,Mhanna2019,Mhanna2017ACD}. 
Approaches for DER coordination typically use prosumer-based decomposition, decomposing the problem at the connection point between prosumers and the network \cite{Scott2019, Andrianesis2019, Attarha2020, Gebbran_SGES}.

In prosumer-based decomposition, the home energy management problem \cite{AZUATALAM2019555} used to schedule prosumer-owned DER is solved in one piece. Because the only information that is exchanged with the aggregator is the grid demand, but not the individual device schedules, this preserves prosumer prerogative and privacy.

Furthermore, because the prosumer subproblems can be solved on single-board computers at prosumer premises \cite{Gebbran_AAMAS}, this approach can be easily implemented in practice, as demonstrated in a recent Australian trial \cite{Scott2019}. 
Existing approaches that use prosumer-based decomposition also solve the network problem in one piece; this results in a significantly lower number of iterations compared to component-based decomposition used, for example, in \cite{Peng2018,Mhanna2019,Mhanna2017ACD}. {\color{black}The total solution time is well below the DER dispatch intervals, which are typically five minutes or more \cite{Scott2019,Gebbran_SGES}. 
Finally, solving the network problem in one piece by the aggregator aligns well with the hierarchical structure of DER aggregation where the aggregator assumes the role of a central coordinator. 
}

\begin{table*}[t]
	\small
	\renewcommand{\arraystretch}{1.2}
	\centering
	\caption{Comparison of the existing approaches for DER coordination and voltage control with the proposed method.}
	\label{table:Methods}
	\begin{tabular}{|l|c|c|c|c|c|c|c|c|c|c|c|}
		\hline
		& \cite{Scott2019,Andrianesis2019,Attarha2020,Gebbran_SGES} & \cite{Farivar_2012} & \cite{Tonkoski2011_TSE,Garcia_2012,Haque_2019,Alyami_2014,Liu_2020} & \cite{Zhou_2020} & \cite{Lusis_2019} &  \cite{DallAnese2014,su_2014,ALMASALMA_2019,Peng2018,Engelmann2019} & This paper \\
		\hline
		\textbf{DOPF with Volt-Var control and fair PV curtailment}  &  &  &  &  &  & & $\times$ \\ 
		DOPF with prosumer-based decomposition & $\times$ &  &  &  &  &  & $\times$ \\ 
		Active \emph{or} reactive inverter dispatch  & & $\times$ & $\times$ &  &  &  &   \\ 
		Active \emph{and} reactive inverter dispatch & $\times$ &   &   & $\times$ & $\times$ & $\times$ & $\times$ \\ 
		Local voltage control with Volt-Var Control  &   &   &   & $\times$ &  &   & $\times$ \\ 
		Fair PV Curtailment & &  & $\times$ &  & $\times$ &   & $\times$ \\   
		\hline
	\end{tabular}
\end{table*}

\subsection{{\color{black}Literature gap}}

{\color{black}Theoretically, ALADIN allows for a prosumer-based decomposition. However, the computation of the sensitivities for the consensus step makes ALADIN impractical for formulations containing nonsmooth functions, such as piecewise-linear functions and problems with integer variables in general. As shown next, these are required in our problem formulation.}
{\color{black}APP, OCD and C+I, on the other hand, do not require a central coordinator, which does not align with the hierarchical prosumer aggregation structure. Further, they require each agent to estimate voltages and power flows in each iteration. This is redundant for prosumer-based approaches, where the aggregator computes the network OPF, obtaining these variables' optima at each iteration, and not just estimates. }
{\color{black}Hence, there are no gains from using OCD or C+I methods in prosumer-based DOPF formulations. Moreover, C+I algorithms have not been applied to non-convex AC OPF \cite{Molzahn2017}.}
{\color{black}Similarly, ATC offers no benefits because the prosumer-based decomposition has only two layers of subproblems, and does not benefit from further partitioning. In case the prosumer-based DOPF was combined with a higher layer OPF (e.g., a medium- and low-voltage coordination problem partitioned in three levels: MV, LV and prosumers), it is possible that ATC could offer benefits, especially if using ADMM to solve parent-child subproblems.}




{\color{black}Given the drawbacks of the existing approaches mentioned above, all the existing prosumer-based DOPF formulations for DER coordination \cite{Scott2019,Andrianesis2019, Gebbran_SGES,Attarha2020} use ADMM. Nevertheless, in the context of Volt-Var Control (VVC) and PV curtailment, ADMM-based approaches} suffer from two main pitfalls. First, they do not consider standard inverter reactive power compensation, which is crucial for voltage regulation in medium- and low-voltage networks \cite{Turitsyn2011}, and is increasingly required by grid connection codes \cite{IEEE_1547-2018, AS_4777-2015,VDE_4105-2018}. 
{\color{black}The standardized VVC curves defined in \cite{IEEE_1547-2018, AS_4777-2015,VDE_4105-2018} are piecewise-linear functions, which adds to the complexity of the already hard, non-convex optimization problem. Therefore, most works that consider reactive power dispatch (in prosumer-based decomposition as well as in other methods) do not account for these standardized functions \cite{Scott2019,Andrianesis2019,Attarha2020,Gebbran_SGES,Lusis_2019,DallAnese2014,su_2014,ALMASALMA_2019,Peng2018}. The only exception is \cite{Zhou_2020}, which uses a control strategy to account for the piecewise-linear VVC in local voltage control instead of using OPF.}

Second, a high PV penetration can result in infeasibility of the OPF problem unless PV generation is curtailed. Still, simply allowing the OPF to curtail PV power, as in \cite{Attarha2020, Gebbran_AAMAS}, puts prosumers at the end of the feeder at a disadvantage \cite{Liu_2020}. In countries like Australia, PV penetration has reached levels that require active network management \cite{OpenEnergyNetworks, Liu_2020}, which calls for a solution that distributes curtailment in an equitable manner. {\color{black}A fair redistribution of PV curtailment has been addressed both from a control perspective \cite{Tonkoski2011_TSE,Garcia_2012,Haque_2019} and from an optimization perspective \cite{Liu_2020,Lusis_2019}. However, both approaches \cite{Liu_2020,Lusis_2019} disregard DER coordination. Moreover, the existing methods use proportional PV curtailment only. We propose two additional PV curtailment methods, using different notions of fairness.}

\subsection{Contributions} \label{Sec:Intro_Contributions}
{\color{black} Table \ref{table:Methods} summarizes the existing approaches for voltage control in low-voltage networks in the context of DER coordination. They treat different aspects of the problem in isolation, but a unified method is still missing. 
Against this background, the major contribution of this paper is a novel OPF formulation for coordination of prosumer-owned DER with a standardized (piecewise-linear) VVC function and a fair PVC, and its implementation in a computationally efficient DOPF. To the best of our knowledge, our formulation is the first to include both a standardized VVC function within an OPF problem, \emph{and} a fair PV curtailment within a DOPF. In a nutshell, this work is a unified distributed framework that incorporates advantages of previous works and further pushes the state-of-the-art for DER coordination algorithms, while avoiding known disadvantages, as shown in Table I.}

In more detail, the proposed method features:
\begin{itemize}
    \item Standard inverter Volt-Var control (VVC) {\color{black}piecewise-linear} function for reactive power compensation embedded in the prosumer subproblem;
    \item Fair PV curtailment (PVC) using \emph{three} different notions of fairness {\color{black}resulting in different curtailment patterns, implemented in the DOPF problem}; 
    \item Tractable\footnote{By ``tractable'' we mean solvable in practice, not polynomial-time computable as in computational complexity theory.} computational performance that scales well owing to its distributed nature {\color{black} despite the mixed-integer nonlinear nature of the problem}.\footnote{In more detail, the problem belongs to the class of mixed-integer nonlinear programming (MINLP) problems that have a non-convex continuous relaxation, which, when compared to MINLP problems with a convex continuous relaxation, are more challenging to solve using current state-of-the-art MINLP technology.}
\end{itemize}

\subsection{Paper structure} \label{Sec:Intro_Structure}
This paper is structured as follows: Section~\ref{Sec:PV_Inverters_VVC} discusses inverter operation modes for voltage control in distribution networks, including the VVC formulation used in the proposed DOPF approach. Section \ref{Sec:DER_Coodrination} formulates the DER coordination problem. Section~\ref{Sec:Fair_PVC} discusses fair PVC strategies using the three proposed equity principles. Section~\ref{Sec:DOPF} details the problem decomposition with PVC redistribution and the resulting formulation of the ADMM-inspired DOPF problem. Section~\ref{Sec:Implementation} discusses the algorithm and implementation details. Section~\ref{Sec:Results} presents detailed results for the 50-bus test system with tree topology, gives a comparison across all 50 simulation cases focusing on the algorithmic performance{\color{black}, and discusses practical implementation considerations}.
Finally, Section~\ref{Sec:Conclusion} gives concluding remarks.

\section{PV inverters for voltage control} \label{Sec:PV_Inverters_VVC}

In traditional distribution networks, voltage regulation only catered for diurnal energy demand fluctuations, which was done at the substation using transformer tap-changers and capacitor banks \cite{Turitsyn2011}. With the emergence of distributed PV generation, the slowly responding utility equipment is no longer fit for purpose. Not only does PV generation significantly reshape net demand patterns, it also causes power fluctuations due to rapid changes in solar insolation, which requires voltage control along the feeder.

PV inverters can address these challenges, and are increasingly being used for feeder voltage control. Various standards \cite{IEEE_1547-2018, AS_4777-2015,VDE_4105-2018} offer different modes of operation: 
\begin{itemize}
    \itemsep0em 
	\item \textbf{Fixed Power Factor} mode (no voltage regulation);
	\item \textbf{Volt-Var Control (VVC)} regulates inverter reactive power based on the voltage at the point of network coupling;
	\item \textbf{Volt-Watt Control} regulates the active power using $P$-$V$ droop control above a certain voltage;
	\item \textbf{Power Rate Limit} mode limits the rate of change of inverter active power output;
	\item\textbf{Voltage Balance} mode balances voltages between phases by injecting unbalanced three-phase current.
\end{itemize}

With the increasing PV penetration, curtailment is required to keep voltages within limits. A network company can impose a limit on the amount of active power that can be injected into the grid, but this is clearly suboptimal. A better solution is to use dynamic curtailment using $P$-$V$ droop control. This can be unfair because the over-voltage at which prosumers are required to curtail PV is not evenly distributed across the grid \cite{Liu_2020}. 
To achieve a fair PV curtailment, {\color{black} different approaches have been proposed, including} using voltage sensitivity information computed centrally \cite{Tonkoski2011_TSE} or in a decentralized fashion \cite{Alyami_2014} {\color{black}to set adaptive droop control for inverters, and using a consensus algorithm \cite{Haque_2019} to coordinate the curtailment control across prosumers. More recently, the authors in \cite{Liu_2020,Lusis_2019} used a centralized DC OPF formulation to compute fair PV curtailment. In \cite{Lusis_2019}, they analyzed the dispatch of optimal active and reactive inverter power considering a fairer curtailment, but without actually distributing all curtailment equally and fairly. This is addressed in \cite{Liu_2020}, which also presents a thorough analysis of proportional and economically fair PVC using different metrics.}

An alternative approach is to cast the voltage control problem as an OPF problem with inverter reactive power compensation and active power curtailment (APC) as decision variables \cite{Farivar_2012,DallAnese2014,su_2014,ALMASALMA_2019}. Authors in \cite{Farivar_2012} and \cite{Yiju_ACM_2020} formulate a relaxed second-order cone (SOC) OPF, where \cite{Farivar_2012} calculates optimal inverter reactive power compensation to minimize line losses, whereas \cite{Yiju_ACM_2020} includes inverter VVC within the OPF to account for reactive compensation when formulating an investment game to assess network PV hosting capacity. 
Other approaches control both reactive power compensation and active power curtailment: \cite{su_2014} formulates a SOC OPF for an unbalanced four-wire distribution network; 
a Jacobi-Proximal ADMM is employed by \cite{ALMASALMA_2019} on a decentralized peer-to-peer network, based on voltage sensitivity coefficients instead of solving the actual OPF problem; and \cite{DallAnese2014} formulates an AC OPF, solving the resulting problem using ADMM to minimize power losses and voltage deviations. 

Yet another approach is proposed in \cite{Zhou_2020}, where the authors study voltage control in distribution networks as a distributed control problem. They show that careful design of local Volt-Var controller is required to ensure convergence. They also demonstrate that the power system dynamics with existing controls can be interpreted as distributed algorithms for solving the optimization problems for reactive power dispatch. The paper, however, does not consider prosumer-owned DER, nor APC.

\begin{figure}[]
    \centering
    \resizebox{!}{3cm}{

\begin{tikzpicture}[x=0.75pt,y=0.75pt,yscale=-1,xscale=1]

\draw  (23,64.83) -- (211.5,64.83)(39.02,13) -- (39.02,118) (204.5,59.83) -- (211.5,64.83) -- (204.5,69.83) (34.02,20) -- (39.02,13) -- (44.02,20)  ;
\draw[line width=0.75pt]    (61.8,31) -- (101,64.7) ;

\draw[line width=0.75pt]    (130,65) -- (168.5,98) ;

\draw[line width=0.75pt]    (101,65) -- (130,65) ;

\draw[line width=0.75pt]    (62,31) -- (39,31) ;

\draw[line width=0.75pt]    (190,98) -- (168.3,98) ;

\draw[thin]  [dash pattern={on 0.84pt off 2.51pt}]  (39.5,98) -- (168.5,98) ;

\draw (65,62) -- ++ (0,5);
\draw (101,62) -- ++ (0,5);
\draw (130,62) -- ++ (0,5);
\draw (168.5,62) -- ++ (0,5);

\draw (65,79) node [scale=0.8]  {$V_{1}$};
\draw (101,79) node [scale=0.8]  {$V_{2}$};
\draw (130,79) node [scale=0.8]  {$V_{3}$};
\draw (168.5,79) node [scale=0.8]  {$V_{4}$};
\draw (27,96) node [scale=0.8]  {$-\overline{q}$};
\draw (33,29) node [scale=0.8]  {$\overline{q}$};
\draw (222,78) node [scale=0.8]  {$V\text{ [V]}$};
\draw (8,16) node [scale=0.8]  {$Q\text{ [VAr]}$};

\end{tikzpicture}}
    \caption{\footnotesize Generic Volt-Var Control (VVC) function.}
    \label{fig:Deadband_VVC}
\end{figure}
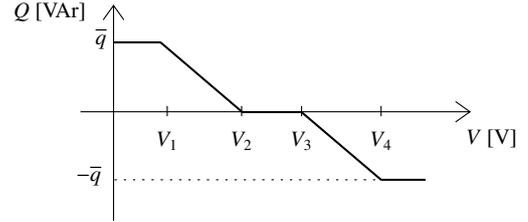

Our work is based upon VVC and APC. 
Unlike \cite{DallAnese2014,Farivar_2012,su_2014} and similar to \cite{Yiju_ACM_2020,Zhou_2020}, our formulation uses a standard VVC function, using a local voltage-based response included in various international standards \cite{IEEE_1547-2018, AS_4777-2015,VDE_4105-2018} and adopted by all manufacturers in Australia \cite{CleanEnergyCouncil}.
VVC is typically represented as a deadband graph, with a region close to the nominal voltage under which no response is enacted. 
The amount of VAr injection or absorption based on voltage and maximum available reactive power at each connection is given by:
%
\begin{equation}
q_{t} = \phi \left( v_t \right) = \begin{cases}
\overline{q}, & \text{if} \quad v_{t} \leq {V}_{1}\\
\overline{q} \frac{ V_{2}-v_{t} }{ V_{2}-V_{1} }, & \text{if} \quad {V}_{1} < v_{t} < {V}_{2}\\
0, & \text{if} \quad {V}_{2} \leq v_{t} \leq {V}_{3}\\
-\overline{q} \frac{ v_{t}-V_{3} }{ V_{4}-V_{3} }, & \text{if} \quad {V}_{3} < v_{t} < {V}_{4}\\
-\overline{q}, & \text{if} \quad v_{t} \geq {V}_{4},
\end{cases} \label{eq:HEM_Reactive}
\vspace{-0mm}
\end{equation}
\normalsize
where $\overline{q}$ is the maximum reactive power output (normally defined as a percentage of the inverter's maximum capacity $\overline{s}$), $v_t$ is the voltage amplitude at the point of common coupling, and ${q}_{t}$ is the reactive power output, at any given time $t$. A graphical representation of \eqref{eq:HEM_Reactive} is shown in Fig.~\ref{fig:Deadband_VVC}.

Different standards suggest or enforce different values for $V_1$ to $V_4$, and implementation follows the requirements of the local distribution network service provider (DNSP) \cite{CleanEnergyCouncil}.

\section{Distributed energy resources coordination} \label{Sec:DER_Coodrination}
We formulate our proposed approach for DER coordination as a multiperiod OPF problem. Conceptually, it consists of two {\color{black}sources of costs. On the one hand}, prosumers \emph{schedule} their DER to minimize energy expenditure\footnote{When ${c}^{\text{ToU}} \! > \! {c}^{\text{FiT}}$, as is the case in Australia, this corresponds to PV self-consumption.}. {\color{black}On the other hand}, the distribution system operator \emph{coordinates} prosumer actions to minimize {\color{black}energy-based operating costs while meeting network and operational constraints. In our formulation, the network cost component is modeled as} the cost of power generation, {\color{black}but it could also be modeled as the marginal cost of losses,} or the cost of drawing power from the grid.

{\color{black}This is conceptually different from bi-level optimization, where the prosumer problem is modeled explicitly, and its solution is included as a constraint in the aggregator’s problem.
Solving bi-level optimization problems requires reformulating the lower-level problem (the prosumer problem in our case) with its optimality KKT conditions, which are then included in the upper level (aggregator) problem, resulting in a MILP problem; however, this can only be done for convex problems \cite{Conejo2020}. Our problem is inherently non-convex due to the AC network constraints and also requires the use of binary variables for the VVC functions, resulting in an MINLP with a non-convex relaxation. Moreover, existing bi-level models (e.g. \cite{Riaz_2019, Zugno_2013}) do not consider network constraints. Solving bi-level models with non-convex mixed-integer problems is an open research problem \cite{Conejo2020}.} 

The objective function {\color{black}of our problem formulation} is
\begin{equation} \label{eq:Central_OPF_1st}
\begin{split}
&\underset{\boldsymbol{x}, \: \boldsymbol{z}} {\mbox{minimize}} \quad F\left( \boldsymbol{x},\boldsymbol{z}\right)  \coloneqq f \big(  \boldsymbol{x} \big)  + \sum_{h \in \mathcal{H}} g_h\left( \boldsymbol{z}_h\right)  \\
& = \underset{t \in \mathcal{T}}{{\sum}} \left( {c}_{2}{ \big(  {p}_{\text{g},0,t}^{+}\big)  }^2 + {c}_{1} {p}_{\text{g},0,t}^{+} + {c}_{0} + \underset{h \in \mathcal{H}}{\sum} \big(  {c}^{\text{ToU}}_{t}{ {p}_{h,t}^{+} } - {c}^{\text{FiT}}{ {p}_{h,t}^{-} \big)  } \right) , 
\end{split}
\end{equation}
\normalsize
where $f \left( \boldsymbol{x}_{} \right) $ represents the network OPF objective function, 
$g_{h} \left( \boldsymbol{z}_{h} \right) $ are prosumer objective functions for each household $h$;
$\mathcal{H}$ is the set of prosumers, 
$\boldsymbol{x}$ is the set of network variables (active and reactive power flows, and voltages, for each $t \in \mathcal{T}$), 
$\boldsymbol{z}_{h}$ is the set of internal variables of prosumer $h$ for each $t \in \mathcal{T}$ (e.g., battery charging and discharging), 
and $\boldsymbol{z} \coloneqq { \left\{ \boldsymbol{z}_h \right\} }_{h\in \mathcal{H}}$. 
The network objective $f \left( \boldsymbol{x}_{} \right)$ can include, for example, loss minimization, peak load reduction, or minimizing the use of backup diesel as in \cite{Scott2019}. 
The prosumer objective $g_{h} \left( \boldsymbol{z}_{h} \right) $ accounts for a fixed time-of-use (ToU) tariff for purchasing energy, and a feed-in-tariff (FiT) for selling excess PV generation to the grid.

For the sake of simplicity, we assume a balanced three-phase network, which can be modeled by its single-phase equivalent. The model can be readily extended to include unbalanced networks with a combination of single- and three-phase connections as in \cite{Scott2019}. 
The network constraints for the OPF are given for each bus $i \in \mathcal{B}$, and for each time interval $t \in \mathcal{T}$ (index 0 denotes the reference bus):
\begin{subequations} \label{eq:OPF_Constraints}
\begin{align}
     & p_{\text{g},i,t} - {p}_{h,t} = v_{i,t} \underset{j \: \in \: \mathcal{B}}{\sum} v_{j,t} \left (g_{ij}\cos\theta_{ij,t}+b_{ij}\sin\theta_{ij,t} \right), \label{eq:OPF_P-flow}\\
    & q_{\text{g},i,t} - {q}_{h,t} = v_{i,t} \underset{j \: \in \: \mathcal{B}}{\sum} v_{j,t} \left (g_{ij}\sin\theta_{ij,t}-b_{ij}\cos\theta_{ij,t} \right), \label{eq:OPF_Q-flow}\\
    %
    & v_{0,t} = 1, \quad \theta_{0,t} = 0, \label{eq:OPF_V-slack}\\
    & \underline{v}_{i} \leq v_{i,t} \leq \overline{v}_{i}, \label{eq:OPF_V-lim}\\
    %
    %
    & \underline{p}_{\text{g},0,t} \leq {p}_{\text{g},0,t} \leq \overline{p}_{\text{g},0,t}, \quad %
    \underline{q}_{\text{g},0,t} \leq {q}_{\text{g},0,t} \leq \overline{q}_{\text{g},0,t}, \label{eq:OPF_G-limits}
\end{align}
\normalsize
\end{subequations}
where $p_{\text{g},i,t}, q_{\text{g},i,t} = 0 \; \forall \; i \in \mathcal{B} \setminus 0$, 
${p}_{h,t}, {q}_{h,t}$ are the total net active and reactive power of prosumer $h$ connected to bus $i$,\footnote{Note that prosumers have their own dedicated buses. In other words, at most one prosumer is connected to at a certain bus $i$, hence $\mathcal{H} \subset \mathcal{B}$.} and $\theta_{ij,t} = \theta_{i,t} - \theta_{j,t}$ is the angle difference between $i$ and its neighboring bus $j$. Constraints \eqref{eq:OPF_P-flow} and \eqref{eq:OPF_Q-flow} model the power flow equations, \eqref{eq:OPF_V-slack} models the reference bus, and \eqref{eq:OPF_V-lim} and \eqref{eq:OPF_G-limits} capture voltage and generator limit constraints.

On the other hand, constraints of prosumers consist of the power balance equation 
%
\begin{equation} \label{eq:HEM_Balance}
{p}_{h,t} = {p}_{h,t}^{\text{bat}} + {p}_{h,t}^{\text{d}} - {p}_{h,t}^{\text{PV}}, \quad \forall \; t \in \mathcal{T}, h \in \mathcal{H},
\end{equation}
\normalsize
where ${p}_{h,t}$ is the total power of household $h$ exchanged with the grid, with $\underline{p}_{h,t} \leq {p}_{h,t} \leq \overline{p}_{h,t}$, ${p}_{h,t}^{\text{bat}}$ is the scheduled battery charging power such that $\underline{p}_{h,t}^{\text{bat}} \leq {p}_{h,t}^{\text{bat}} \leq \overline{p}_{h,t}^{\text{bat}}$, ${p}_{h,t}^{\text{d}}$ is the household fixed demand (non-controllable), and ${p}_{h,t}^{\text{PV}}$ is the PV generation power output. Additionally, let ${p}_{h,t} = {p}_{h,t}^{+} - {p}_{h,t}^{-}$, where ${p}_{h,t}^{+}$ and ${p}_{h,t}^{-}$ are non-negative terms that represent the imported and exported active power.\footnote{Note that because the second term in \eqref{eq:Central_OPF_1st} is a convex piecewise linear function, \emph{at least one} of the variables ${p}_{h,t}^{+}$ and ${p}_{h,t}^{-}$ can be zero at time slot $t$. This therefore obviates the need to use binary variables.
}

Now, we enable a reduction in PV generation by introducing a schedulable curtailment variable $y_{h,t}$. 
The inverter's capacity is also considered, which may require additional active power curtailment according to the reactive power response. 
The following equations are introduced $ \forall \: t \in \mathcal{T}, h \in \mathcal{H}$:
%
\begin{subequations} \label{eq:HEM_PVC}
\begin{align}
& 0 \leq {p}_{h,t}^{\text{PV}} \leq \tilde{p}_{h,t}^{\text{PV}}, \label{eq:HEM_PV_lim} \\
& {y}_{h,t} = \tilde{p}_{h,t}^{\text{PV}} - {p}_{h,t}^{\text{PV}}, \label{eq:HEM_PVC_Gen} \\
& \overline{s}_{h}^{2} \geq { \left( {p}_{h,t}^{\text{PV}} \right) }^2 + {q}^{2}_{h,t}, \label{eq:HEM_Inv_Cap}
\end{align}
\end{subequations}
\normalsize
where $\tilde{p}_{h,t}^{\text{PV}}$ is a parameter denoting the maximum PV generation for prosumer $h$ at $t$, and ${y}_{h,t} \geq 0$ denotes the amount of curtailed PV power.\footnote{In our model, we assume batteries and PV have independent inverters. A hybrid inverter would require slight modifications in (\ref{eq:HEM_Inv_Cap}).} For batteries, the constraints are, $ \forall \: t \in \mathcal{T}, h \in \mathcal{H}$:
%
\begin{subequations} \label{eq:HEM_Bat}
\begin{align}
& {p}_{h,t}^{\text{bat}} = {p}_{h,t}^{\text{ch}} - {p}_{h,t}^{\text{dis}}, \label{eq:HEM_bat_Dec}\\
& {SoC}_{h,0} \le {SoC}_{h,T}, \label{eq:HEM_SoC0}\\
& {SoC}_{h,t} = {SoC}_{h, t - \Delta t } + \left( \eta_{h}^{\text{ch}} {p}_{h,t}^{\text{ch}} - {{p}_{h,t}^{\text{dis}}}/{\eta_{h}^{\text{dis}}} \right)\Delta t, \label{eq:HEM_bat_SoC}
\end{align}
\end{subequations}
\normalsize
where ${p}_{h,t}^{\text{ch}}, {p}_{h,t}^{\text{dis}} \geq 0$ are the battery charging and discharging powers respectively, ${SoC}_{h,t}$ is the scheduled battery state of charge, with $\underline{SoC}_{h,t} \leq {SoC}_{h,t} \leq \overline{SoC}_{h,t},$\footnote{The inclusion of \eqref{eq:HEM_SoC0} avoids the full depletion of the battery, without considering the subsequent time horizon. Replacing this constraint would be recommended when implementing the algorithm in a rolling horizon basis.} $\eta_{h}$ is the battery charge or discharge efficiency, and $\Delta t$ is the time interval.

Let OPF constraints \eqref{eq:OPF_Constraints} define a feasible set $\mathcal{X}$ for network variables $\boldsymbol{x}$. Likewise, let prosumer constraints \eqref{eq:HEM_Balance}, \eqref{eq:HEM_PVC}, and \eqref{eq:HEM_Bat} define a feasible set $\mathcal{Z}_{h}$ for the variables $\boldsymbol{z}_h$ (prosumer's active and reactive power demand, battery charging, state of charge, and PV curtailment). Henceforth, $\boldsymbol{x} \in \mathcal{X}$ and $\boldsymbol{z}_h \in \mathcal{Z}_{h}$, and $\mathcal{Z}$ is the feasible set for all prosumer variables $\boldsymbol{z}$. 

We can rewrite the above problem in its compact form, accounting for the VVC coupling between prosumers and the network, which yield the following mixed-integer nonlinear programming (MINLP) problem:\footnote{As no two prosumers share the same bus and $\mathcal{H} \subset \mathcal{B}$, as discussed previously, we have $v_{h,t} = v_{i,t}$ for prosumer $h$ connected at bus $i$.}
%
\begin{subequations} \label{eq:Central_OPF_2nd}
\begin{align}
& \underset{\boldsymbol{x} \in \mathcal{X}, \: \boldsymbol{z} \in \mathcal{Z}} {\mbox{minimize}} \quad F \left( \boldsymbol{x},\boldsymbol{z} \right) \label{eq:Central_OPF_2nd_Obj} \\
& \hspace{0.01cm} \text{subject to} \hspace{0.27cm} q_{h,t} = \phi \left( v_{h,t} \right) , \quad \forall \; h \in \mathcal{H}, \; t \in \mathcal{T}. \label{eq:Central_OPF_2nd_VVC}%
\end{align}
\normalsize
\end{subequations}
%
%

This DER coordination problem is an all-encompassing model that can be used to compute a DER-based OPF with VVC, either centrally or in a distributed fashion. Previous works have utilized similar problems, but \cite{Attarha2020} and \cite{Gebbran_AAMAS} do not consider the VVC given by (\ref{eq:Central_OPF_2nd_VVC}), whereas \cite{Scott2019} and \cite{Andrianesis2019} do not account for the VVC nor the PVC given by (\ref{eq:HEM_PVC_Gen}), and \cite{Yiju_ACM_2020} accounts for the VVC in a SOC OPF but does not consider schedulable DER. 
However, this problem still lacks consideration of the fair distribution of PV curtailment across prosumers, which we discuss next. 

\section{Fair PV curtailment redistribution} \label{Sec:Fair_PVC}
To control the amount of power curtailed at prosumer PV generation, a penalty function is added to \eqref{eq:Central_OPF_2nd_Obj}: 
\begin{subequations} \label{eq:Central_OPF_PVC}
\begin{align}
&\underset{\boldsymbol{x} \in \mathcal{X}, \: \boldsymbol{z} \in \mathcal{Z}} {\mbox{minimize}} \quad  F \left( \boldsymbol{x},\boldsymbol{z} \right) + \alpha \Psi \text{(} . \text{)}, \label{eq:Central_OPF_PVC_Obj}\\
& \hspace{0.01cm} \text{subject to} \hspace{0.27cm} q_{h,t} = \phi \left( v_{h,t} \right) , \quad \forall \; h \in \mathcal{H}, \; t \in \mathcal{T} \label{eq:Central_OPF_PVC_VVC},%
\end{align}
\normalsize
\end{subequations}
\noindent where $\alpha > 0$ is a numerical weight that determines how much a reduction in $\Psi \text{(} . \text{)}$ is favored over the cost minimization $F \left( \boldsymbol{x},\boldsymbol{z} \right) $ or vice versa. For $\alpha=0$ we recover \eqref{eq:Central_OPF_2nd}. 

Next, constraints are introduced to problem \eqref{eq:Central_OPF_PVC} to model the type of curtailment, encoding different equity principles: egalitarian, proportional, and uniform dynamic. 

\subsection{Egalitarian PVC redistribution} \label{Sec:Fair_PVC_Ega}
In this case, PVC is redistributed equally among all prosumers. An example of this case is shown in Fig.~\ref{fig:PVC_Redistribution}b). It can be mathematically translated into an egalitarian objective, where the end-goal is to achieve the lowest PVC across all prosumers. The constraint to be included in \eqref{eq:Central_OPF_PVC} is:
%
\begin{equation} \label{eq:PVC_case1}
\overline{y}_{t} \geq {y}_{h,t}, \qquad \forall \; h \in \mathcal{H}, \; t \in \mathcal{T},
\end{equation}
\normalsize
where $\overline{y}_t$ is the highest amount of curtailed PV ${y}_{h,t}$ at any given house, at time period $t$. Correspondingly, we use $\Psi \left( \boldsymbol{\overline{y}} \right) = \sum_{t} \overline{y}_{t}$ as the penalty function.


\subsection{Proportional PVC redistribution} \label{Sec:Fair_PVC_Prop}
In this case, each prosumer is curtailed proportionally to how much they would be able to export on an uncongested operation (which is different from their actual PV generation).\footnote{Note that it is also possible to determine a proportional PVC based on installed PV capacities \cite{Liu_2020}. However, the self-interested allocation of PV has been shown to be inefficient \cite{Yiju_ACM_2020}.} This is shown in Fig.~\ref{fig:PVC_Redistribution}c). Once again, we use $\Psi \left( \boldsymbol{\overline{y}} \right) = \sum_{t} \overline{y}_{t}$, but for the proportional PVC redistribution, the corresponding constraint to be considered in \eqref{eq:Central_OPF_PVC} is:
%
\begin{equation} \label{eq:PVC_case2_const}
    \overline{y}_{t} \geq \frac{{y}_{h,t}}{{p}_{h,t}^{-} + {y}_{h,t}}, \qquad \forall \; h \in \mathcal{H}, \; t \in \mathcal{T},
\end{equation}
\normalsize
where ${p}_{h,t}^{-}$ is the exported power of house $h$ during period $t$. A proportional factor now divides ${y}_{h,t}$, corresponding to the maximum available export power of each prosumer. 

\subsection{Uniform dynamic PVC redistribution} \label{Sec:Fair_PVC_UD}
\begin{figure}[!t]
    \centering
    \resizebox{!}{3.6cm}{\input{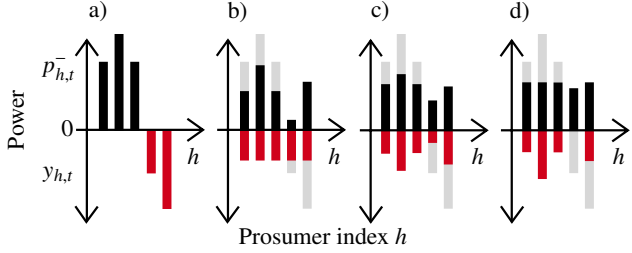}}
    \caption{Examples of PVC redistribution for five households where black indicates power exported and red power curtailed: a) shows the unfair base case (i.e., curtailed as in \cite{Attarha2020, Gebbran_AAMAS}) where three prosumers only export energy (black), and two prosumers are completely curtailed (red), b) shows an egalitarian redistribution, c) depicts the proportional redistribution, and d) demonstrates the uniform dynamic PVC redistribution. The gray bars show the base case. Note the solid colors retain the same total length in all cases.\\
    \vspace{5mm}}
    \label{fig:PVC_Redistribution}
\end{figure}

This case employs a uniform dynamic PVC redistribution that reduces the highest power-exporting prosumers first. Observe in Fig.~\ref{fig:PVC_Redistribution}d) that the optimisation algorithms enforces a smallest possible uniform export limit. The limit is determined based on the current operating condition, hence the name \textit{dynamic}. This stands in contrast to the current industry practice of imposing a fixed export limit irrespective of the operating condition. We use $\Psi \left(\underline {\boldsymbol{y}} \right) = \sum_{t} \left( - \underline{y}_{t} \right)$. Its associated constraint is: 
%
\begin{equation} \label{eq:PVC_case3}
\underline{y}_{t} \leq {p}^{{-}}_{h,t}, \quad \forall \; { \left\{ h \in \mathcal{H} \mid \left( {p}^{{-}}_{h,t}+{y}_{h,t} \right) > \underline{y}_{t} \right\} }, \; t \in \mathcal{T},
\end{equation}
\normalsize
where $\underline{y}_t$ is the lowest amount of exported power ${p}^{{-}}_{h,t}$ at any given house subject to curtailment, at time period $t$. In other words, prosumers exporting less than $\underline{y}_t$ are not curtailed.

\section{Distributed coordination problem} \label{Sec:DOPF}
The resulting DER coordination problem \eqref{eq:Central_OPF_PVC} and the corresponding PVC constraint \eqref{eq:PVC_case1}-\eqref{eq:PVC_case3}, account for a fair PVC sharing among prosumers on top of the reactive power compensation from the VVC. 

Three main complications arise when solving this problem centrally. First, the privacy of all prosumers is violated because prosumers have to share their energy consumption details with the central coordinator. Second, this problem is difficult to solve, consisting of a large mixed-integer problem with a non-convex continuous relaxation that is computationally challenging in itself \cite{Verma2019}. Third, the function introduced by the VVC function \eqref{eq:HEM_Reactive} is piecewise linear (PWL) and non-convex. By including it in the OPF formulation \eqref{eq:Central_OPF_PVC_VVC}, the problem becomes an MINLP with a number of binary variables proportional to $HT$.

The problem in its central form \eqref{eq:Central_OPF_PVC} cannot be solved in a distributed fashion because it is not directly decomposable. Two steps are necessary to decouple common variables between the aggregator and prosumers in order to bestow a decomposable structure, which gives rise to $1+H$ subproblems.

\subsection{Problem decomposition} \label{Sec:DOPF_Decomposition}

First, constraints associated with VVC \eqref{eq:Central_OPF_PVC_VVC} and PVC \eqref{eq:PVC_case1}-\eqref{eq:PVC_case3} need to be included in either the network feasible set $\mathcal{X}$ or the prosumer feasible sets $\mathcal{Z}$. As the PVC constraint is associated with the PVC penalty function, it is included in the network set $\mathcal{X}$. Now, the new feasible set for the network is $\hat{\mathcal{X}}$. Similarly, as the VVC-related constraint is related to the inverter's reactive power output, we include this equation in the prosumer sets $\mathcal{Z}$, which then becomes $\hat{\mathcal{Z}}$. Another reason for this attribution of the VVC equation to prosumer sets is to avoid further complicating the already NP-hard network subproblem by adding the mixed-integer variables associated with the PWL function.

Second, since some variables appear in both $\hat{\mathcal{X}}$ and $\hat{\mathcal{Z}}$, the general approach in ADMM-based solutions is to duplicate these variables by assigning one copy for the network and one copy for the respective prosumer. Next, consensus between these copies is captured by, $\forall \ h \in \mathcal{H}, \ t \in \mathcal{T}$,
%
\begin{subequations} \label{eq:Coupling}
\begin{align}
\hat{p}_{h,t} &= {p}_{h,t}, \label{eq:Coupling_p}\\
\hat{y}_{h,t} &= {y}_{h,t}, \label{eq:Coupling_PVC}\\
\hat{q}_{h,t} &= {q}_{h,t}, \label{eq:Coupling_q}\\
{v}_{i,t} &= {v}_{h,t}, \: { \Big{\{} i \in \mathcal{B} \mid \exists \; h \; \text{connected to} \; i \Big{\}} }, \label{eq:Coupling_V}%
\end{align} 
\normalsize
\end{subequations}
where the left-hand terms are copies for the network problem, $\left\{ \hat{p}_{h,t}, \hat{y}_{h,t}, \hat{q}_{h,t}, {v}_{i,t} \right\} \subset {\boldsymbol{\hat{x}}} \in \hat{\mathcal{X}}$, and the right-hand terms are copies for the prosumer problem, $\left\{ {p}_{h,t}, {y}_{h,t}, {q}_{h,t}, {v}_{h,t} \right\} \subset \hat{\boldsymbol{z}} \in \hat{\mathcal{Z}}$. This confers a decomposable structure on the problem. 
In more detail, duplicating the common variables allows us to rewrite problem \eqref{eq:Central_OPF_PVC} as
%
\begin{subequations} \label{eq:Decomposed_OPF}
\begin{align}
&\underset{\boldsymbol{\hat{x}} \: \in \: \mathcal{\hat{X}}, \: \boldsymbol{\hat{z}} \: \in \: \hat{\mathcal{Z}}} {\mbox{minimize}} 
\quad F \left( \boldsymbol{\hat{x}},\boldsymbol{\hat{z}} \right) + \alpha \Psi \text{(} . \text{)}, \label{eq:Decomposed_OPF_Obj}\\
%
& \hspace{0.01cm} \text{subject to} \hspace{1.30cm} \text{(\ref{eq:Coupling})} \label{eq:Decomposed_OPF_const},%
\end{align}
\end{subequations}
\normalsize
where $\boldsymbol{\hat{x}} \in \hat{\mathcal{X}}$ is the new set of network variables, and $\boldsymbol{\hat{z}} \in \hat{\mathcal{Z}}$ is the new set of prosumer variables. 
%
%
In this form, the sets of variables $\mathcal{\hat{X}}$ and $\hat{\mathcal{Z}}$ are decoupled, and \eqref{eq:Decomposed_OPF_Obj} is separable if \eqref{eq:Decomposed_OPF_const} is relaxed (shown next).
We will exploit this decomposable structure to solve the resulting problem in a distributed fashion. 

%
%
The ADMM \cite{Boyd2011} capitalises on the decomposable structure in \eqref{eq:Decomposed_OPF} and performs alternating minimizations over sets $\mathcal{\hat{X}}$ and $\hat{\mathcal{Z}}_{h}$. At each iteration $k$, each subproblem calculates the next set of variables denoted by $k+1$. The prosumer subproblems utilize the next set of variables $k+1$ calculated by the network OPF subproblem. 

%
To begin with, we write the Augmented (partial) Lagrange function of the problem:
\begin{equation} \label{eq:Lagrangian}
\small
\begin{split}
& L \coloneqq F \left( \boldsymbol{\hat{x}},\boldsymbol{\hat{z}} \right) + \alpha \Psi \text{(} . \text{)} + \underset{t \in \mathcal{T}}{\sum} \underset{h \in \mathcal{H}}{\sum} \Big( \frac{\rho^{\text{p}}}{2} \left( \hat{p}_{h,t} - {p}_{h,t} \right)^2 \\
& \hspace{3mm} + \lambda^{\text{p}}_{h,t} \left( \hat{p}_{h,t} - {p}_{h,t} \right) + \frac{\rho^{\text{y}}}{2} \left( \hat{y}_{h,t} - {y}_{h,t} \right)^2 + \lambda^{\text{y}}_{h,t} \left( \hat{y}_{h,t} - {y}_{h,t} \right) \\
& \hspace{3mm} + \frac{\rho^{\text{q}}}{2} \left( \hat{q}_{h,t} - {q}_{h,t} \right)^2 + \lambda^{\text{q}}_{h,t} \left( \hat{q}_{h,t} - {q}_{h,t} \right) \Big) \\
& \hspace{3mm} = F \left( \boldsymbol{\hat{x}},\boldsymbol{\hat{z}} \right) +  \alpha \Psi \text{(} . \text{)} +\underset{h \in \mathcal{H}}{\sum} \big( L^{\text{p}}_{h} + L^{\text{y}}_{h} + L^{\text{q}}_{h} \big) \\
& \hspace{3mm} = F \left( \boldsymbol{\hat{x}},\boldsymbol{\hat{z}} \right) +  \alpha \Psi \text{(} . \text{)} + \underset{h \in \mathcal{H}}{\sum} L_{h}, 
\normalsize
\end{split}
\end{equation}
where $\rho > 0$ is a penalty parameter associated with each type of variable, and ${\lambda}$ is the dual variable (Lagrangian multiplier) associated with each coupling constraint.\footnote{The dual variables for active power are used as locational marginal prices (LMP) in other approaches to the DER coordination problem. See \cite{Papavasiliou_2018} for a discussion on LMP.} 

Before we reach the final ADMM-inspired formulation, we need to discuss three additional steps due to the nature of the problem.

\subsection{VVC relaxation} \label{Sec:DOPF_VVC}
%
The voltage at prosumer $v_{h,t}$ is not determined in any way by the household, but calculated in the network subproblem. 
Therefore, instead of taking the Augmented Lagrangian of the voltage, we instead 
(i) treat $v^{k+1}_{h,t}$ as an input to the prosumer subproblems, which yields $q^{k+1}_{h,t}$ with \eqref{eq:HEM_Reactive},
and (ii) include a modified penalty term to the network Lagrangian.

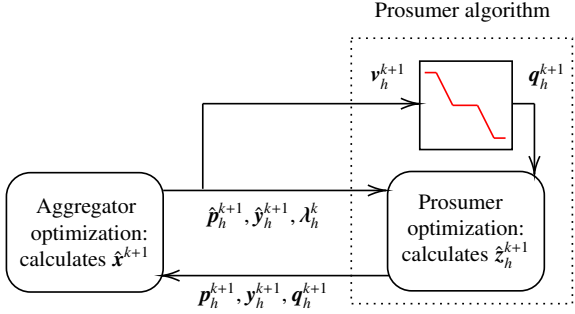
\begin{figure}[!tp]
    \centering
    \resizebox{!}{4.25cm}{\tikzset{every picture/.style={line width=0.75pt}} 

\begin{tikzpicture}[x=0.75pt,y=0.75pt,yscale=-1,xscale=1]

\draw   (9.5,149.57) .. controls (9.5,141.53) and (16.03,135) .. (24.08,135) -- (89.93,135) .. controls (97.97,135) and (104.5,141.53) .. (104.5,149.57) -- (104.5,193.3) .. controls (104.5,201.35) and (97.97,207.88) .. (89.93,207.88) -- (24.08,207.88) .. controls (16.03,207.88) and (9.5,201.35) .. (9.5,193.3) -- cycle ;
\draw    (128.5,92.86) -- (257.5,92.86) ;
\draw [shift={(259.5,92.86)}, rotate = 180] [color={rgb, 255:red, 0; green, 0; blue, 0 }  ][line width=0.75]    (10.93,-3.29) .. controls (6.95,-1.4) and (3.31,-0.3) .. (0,0) .. controls (3.31,0.3) and (6.95,1.4) .. (10.93,3.29)   ;
\draw    (104.5,145.86) -- (237.5,145.86) ;
\draw [shift={(239.5,145.86)}, rotate = 180] [color={rgb, 255:red, 0; green, 0; blue, 0 }  ][line width=0.75]    (10.93,-3.29) .. controls (6.95,-1.4) and (3.31,-0.3) .. (0,0) .. controls (3.31,0.3) and (6.95,1.4) .. (10.93,3.29)   ;
\draw    (241.5,197.86) -- (105.5,197.86) ;
\draw [shift={(103.5,197.86)}, rotate = 360] [color={rgb, 255:red, 0; green, 0; blue, 0 }  ][line width=0.75]    (10.93,-3.29) .. controls (6.95,-1.4) and (3.31,-0.3) .. (0,0) .. controls (3.31,0.3) and (6.95,1.4) .. (10.93,3.29)   ;
\draw    (329.5,92.86) -- (329.5,133.85) ;
\draw [shift={(329.5,135.85)}, rotate = 270] [color={rgb, 255:red, 0; green, 0; blue, 0 }  ][line width=0.75]    (10.93,-3.29) .. controls (6.95,-1.4) and (3.31,-0.3) .. (0,0) .. controls (3.31,0.3) and (6.95,1.4) .. (10.93,3.29)   ;
\draw    (316.5,92.86) -- (329.5,92.86) ;
\draw    (128.5,92.86) -- (128.5,144.86) ;
\draw [color={rgb, 255:red, 253; green, 0; blue, 0 }  ,draw opacity=1 ]   (270,73.88) -- (280,93.88) ;
\draw [color={rgb, 255:red, 253; green, 0; blue, 0 }  ,draw opacity=1 ]   (295,93.88) -- (305,113.88) ;
\draw [color={rgb, 255:red, 253; green, 0; blue, 0 }  ,draw opacity=1 ]   (280,93.88) -- (295,93.88) ;
\draw [color={rgb, 255:red, 253; green, 0; blue, 0 }  ,draw opacity=1 ]   (305,113.88) -- (312,113.88) ;
\draw [color={rgb, 255:red, 253; green, 0; blue, 0 }  ,draw opacity=1 ]   (263,73.88) -- (270,73.88) ;
\draw   (260,66.88) -- (316,66.88) -- (316,120.88) -- (260,120.88) -- cycle ;
\draw  [dash pattern={on 0.84pt off 2.51pt}] (218.5,52.88) -- (352.5,52.88) -- (352.5,215) -- (218.5,215) -- cycle ;
\draw   (240.5,148.86) .. controls (240.5,140.81) and (247.03,134.28) .. (255.08,134.28) -- (320.93,134.28) .. controls (328.97,134.28) and (335.5,140.81) .. (335.5,148.86) -- (335.5,192.58) .. controls (335.5,200.63) and (328.97,207.16) .. (320.93,207.16) -- (255.08,207.16) .. controls (247.03,207.16) and (240.5,200.63) .. (240.5,192.58) -- cycle ;

\draw (288,170.72) node   [align=left] { \ \ \ Prosumer\\ \ optimization: \\calculates $\displaystyle \hat{\boldsymbol{z}}^{k+1}_{h}$};
\draw (165,161) node    {$\hat{\boldsymbol{p}}^{k+1}_{h} ,\hat{\boldsymbol{y}}^{k+1}_{h} ,\boldsymbol{\lambda }^{k}_{h}$};
\draw (241,76) node    {$\textcolor[rgb]{0,0,0}{\boldsymbol{v}}\textcolor[rgb]{0,0,0}{^{k+1}_{h}}$};
\draw (167,210) node    {$\boldsymbol{p}^{k+1}_{h} ,\boldsymbol{y}^{k+1}_{h} ,\boldsymbol{q}^{k+1}_{h} \ $};
\draw (337,76) node    {$\textcolor[rgb]{0,0,0}{\boldsymbol{q}}\textcolor[rgb]{0,0,0}{^{k+1}_{h}}$};
\draw (286,37) node   [align=left] {Prosumer algorithm};
\draw (57,171.44) node   [align=left] { \ \ \ Aggregator \\ \ \ optimization: \\calculates $\displaystyle \hat{\boldsymbol{x}}^{k+1}$};

\end{tikzpicture}}
    \caption{\footnotesize Algorithm flowchart for DOPF with prosumer-based VVC.\\ 
    \vspace{7mm}}
    \label{fig:Flowchart_VVC}
\end{figure}

From the first point, the PWL function for the VVC \eqref{eq:HEM_Reactive} is no longer part of the prosumer optimization subproblem. Instead, it is calculated externally, as is shown in Fig.~\ref{fig:Flowchart_VVC}. As a result, $q^{k+1}_{h,t}$ becomes a parameter in the prosumer subproblem. We define for each household $\Tilde{L}_h = L^{\text{p}}_{h}+L^{\text{y}}_{h}$. 

From the second point, we introduce a new penalty parameter that damps oscillations of ${v}_{i,t}$ in subsequent iterations: 
%
\begin{equation}
    \Phi \left( \boldsymbol{\hat{q}} \right) = \beta \underset{t \in \mathcal{T}}{\sum} \underset{h \in \mathcal{H}}{\sum} { \big( \hat{q}_{h,t} - \hat{q}_{h,t}^{k} \big) }^2, \label{eq:OPF_Obj_VVC}\\
\end{equation}
\normalsize
where $\beta$ is a weighting factor. This equation penalizes oscillations of reactive power $\hat{q}^{k+1}_{h,t}$, which occur in ill-conditioned problems. This can lead to a violation of constraint \eqref{eq:Coupling_V} that does not appear in the Lagrangian, because it is not a variable in the prosumer subproblems.

This procedure greatly simplifies the network problem by removing the binary variables associated with the non-convex piece-wise linear VVC function.

\subsection{PVC constraints} \label{Sec:DOPF_PVC}
Since \eqref{eq:HEM_PVC_Gen} is not accounted for in the network subproblem, we need to limit the amount of curtailed power. We do so by adding the following constraint to $\hat{\mathcal{X}}$: 
%
\begin{equation} \label{eq:PVC_ADMM_2}
    {p}^{-}_{h,t} + \hat{y}_{h,t} \leq {p}^{-,k}_{h,t} + {y}_{h,t}^{k},  \quad \forall \; h \in \mathcal{H}, \; t \in \mathcal{T},
\end{equation}
\normalsize
which guarantees the sum of exported and curtailed power on the aggregator problem is never larger than the present sum of the exported and curtailed power. Therefore, if a prosumer is generating power but also consuming the same amount of power within its household, it will not be curtailed.

Moreover, a modification is required for the proportional PVC redistribution constraint; the prosumer constraints that model the power and curtailed power are no longer within the OPF (network) subproblem. Therefore, \eqref{eq:PVC_case2_const} is rewritten using prosumer values from the previous iteration as parameters. We have $\forall \; h \in \mathcal{H}, \; t \in \mathcal{T}$:
%
\begin{equation} \label{eq:PVC_ADMM_3}
    {\overline{y}_{t}} \geq \frac{\underset{h \in \mathcal{H}}{\sum} {y}_{h,t}^{k}} {\left( {p}_{h,t}^{-,k} + {y}_{h,t}^{k} + \gamma \right) \underset{h \in \mathcal{H}}{\sum} \left( {p}_{h,t}^{-,k} + {y}_{h,t}^{k} \right) } {{\hat{y}_{h,t}}},
\end{equation}
\normalsize
where $\gamma > 0$ is a very small value to prevent an undefined operation if no power is exported or curtailed. 


\subsection{Power flow equations} \label{Sec:DOPF_PF_Eq}
The curtailed power value is changed within the network subproblem, so a small modification of the power flow equation is needed to capture the resulting difference in each prosumer's power demand.
Hence, we update the left-hand side of the OPF constraint \eqref{eq:OPF_P-flow}, where $\forall \: i \in \mathcal{B}$:
%
\begin{equation} \label{eq:OPF_const_PF_updated}
    \begin{split}
    & {p}_{\text{g},i,t} - \left( \hat{p}_{h,t} + \big( \hat{y}_{h,t} - {y}_{h,t}^{k} \big) \right) \\
    & = v_{i,t} \underset{j \: \in \: \mathcal{B}}{\sum} v_{j,t} \left( g_{ij}\cos\theta_{ij,t}+b_{ij}\sin \theta_{ij,t} \right),
    \end{split}
\end{equation}
\normalsize
which reflects the new power demand from each household; that is, considering the changes within the PVC variable in addition to the updated variable for prosumer's demand $\hat{p}_{h,t}$.

\subsection{Resulting ADMM-inspired algorithm} \label{Sec:DOPF_ADMM}
 
Finally, at a given iteration $k$, the next iterate $k+1$ is computed by solving the following subproblems for primal variables:
%
\begin{subequations} \label{eq:ADMM}
\begin{align}
    & \boldsymbol{\hat{x}}^{k+1} \coloneqq \underset{\boldsymbol{\hat{x}} \: \in \: \mathcal{\hat{X}}}{\mbox{argmin}} \; \big[ F \left( \boldsymbol{\hat{x}},\boldsymbol{\hat{z}} \right) + \Phi \left( \boldsymbol{\hat{q}} \right) + \alpha \Psi \text{(} . \text{)} + \underset{h \in \mathcal{H}}{\sum} L_{h} \big], \label{eq:xk+1}\\
    & {q}^{k+1}_{h,t} \coloneqq \phi \left( {v}^{k+1}_{h,t} \right), \hspace{1.2cm} \forall \; h \in \mathcal{H}, \; t \in \mathcal{T},\label{eq:qk+1} \\
    & \boldsymbol{\hat{z}}^{k+1}_{h} \coloneqq \underset{\hat{\boldsymbol{z}}_{h} \: \in \: \hat{\mathcal{Z}}_{h}}{\mbox{argmin}} \; \big[ g \left( {\boldsymbol{\hat{z}}}_{h} \right) + \Tilde{L}_h \big] \hspace{6.2mm} \forall \; h \in \mathcal{H}, \label{eq:zk+1}\\
    \intertext{ \normalsize {and update dual variables $\forall \; h \in \mathcal{H}, \; t \in \mathcal{T}$:}} 
    & {\lambda}^{\text{p},k+1}_{h,t} \coloneqq {\lambda}^{\text{p},k}_{h,t} + \rho^{\text{p}} \left( {\hat{p}_{h,t}}^{k+1} - {p}_{h,t}^{k+1} \right) ,\label{eq:Lk+1} \\
    & {\lambda}^{\text{y},k+1}_{h,t} \coloneqq {\lambda}^{\text{y},k}_{h,t} + \rho^{\text{y}} \left( {\hat{y}_{h,t}}^{k+1} - {y}_{h,t}^{k+1} \right) ,\label{eq:L1k+1} \\
    & {\lambda}^{\text{q},k+1}_{h,t} \coloneqq {\lambda}^{\text{q},k}_{h,t} + \rho^{\text{q}} \left( {\hat{q}_{h,t}}^{k+1} - {q}_{h,t}^{k+1} \right) ,\label{eq:L2k+1}
    %
    %
\end{align}
\end{subequations}
\normalsize
where \eqref{eq:xk+1} is the subproblem solved by the aggregator (holding $\boldsymbol{p},\boldsymbol{y}, \boldsymbol{q}$ constant at $k$), \eqref{eq:qk+1} is the VVC update, \eqref{eq:zk+1} is the subproblem of each prosumer (holding $\hat{\boldsymbol{p}},\hat{\boldsymbol{y}}, \hat{\boldsymbol{q}}$ constant at $k+1$, results of the network subproblem and VVC update), and \eqref{eq:Lk+1}-\eqref{eq:L2k+1} are, respectively, the dual updates for $\boldsymbol{\lambda}^{\text{p}}, \boldsymbol{\lambda}^{\text{y}}$ and $\boldsymbol{\lambda}^{\text{q}}$. 
Because problems \eqref{eq:qk+1} and \eqref{eq:zk+1} are decoupled, they can be solved in parallel, for example, by individual prosumers. 
\section{Implementation} \label{Sec:Implementation}
The proposed coordination method is described in Algorithm 1. The problem \eqref{eq:ADMM} is solved on a $\SI{32}{\giga\byte}$ RAM, Intel i7-7700, $\SI{3.60}{\giga\hertz}$ PC. All subproblems were implemented in Python using Pyomo \cite{Pyomo2011Hart} as a modeling interface, and solved using IPOPT 
v3.12.11 \cite{IPOPT}, with linear solver MA27 \cite{HSL1}. All subproblems were solved sequentially. 
%
\begin{figure}[!t]
%
\begin{algorithm}[H]
  \caption{DOPF with VVC and PVC Redistribution}
  \begin{algorithmic}[1]
    \STATE {\textbf{Initialization:} $k=1, \boldsymbol{\lambda}^{1} = 0, \boldsymbol{\rho}^{1} = 1$}, $v_{h,t} = v_{i,t} = 1$  p.u., all other variables set to zero.
    \WHILE{${\lVert\boldsymbol{r}^{k}\rVert}_2 > \epsilon^{\text{pri}}$ \textbf{or} ${\lVert\boldsymbol{s}^{k}\rVert}_2 > \epsilon^{\text{dual}}$}
        \STATE Compute $\boldsymbol{\hat{x}}^{k+1}$ using (\ref{eq:xk+1}) at aggregator.
        \STATE Communicate $\boldsymbol{\hat{p}}_{h}^{k+1}, \boldsymbol{\hat{y}}_{h}^{k+1}, \boldsymbol{v}_{h}^{k+1}, \boldsymbol{\lambda}^{k}_{h}$ to all prosumers.
        \STATE Calculate (\ref{eq:qk+1}), in parallel at each prosumer
        \STATE Compute $\boldsymbol{\hat{z}}^{k+1}$ using (\ref{eq:zk+1}), in parallel at each prosumer
        \STATE Communicate all $\boldsymbol{p}_{h}^{k+1}, \boldsymbol{y}_{h}^{k+1}, \boldsymbol{q}^{k+1}_{h}$ to aggregator
        \STATE Update $\boldsymbol{\lambda}^{k+1}$ using (\ref{eq:Lk+1})-(\ref{eq:L2k+1})
        \STATE Update $\boldsymbol{\rho}^{k+1}$ using  (\ref{eq:ADMM_mu})
        \STATE $k = k + 1$
    \ENDWHILE
  \end{algorithmic}
\end{algorithm}
\end{figure}

\subsection{ADMM algorithmic details} \label{Sec:Implementation-ADMM}
We use primal and dual residuals to define the stopping criteria \cite{Boyd2011}. They are, respectively:
%
\begin{subequations} \label{eq:Residuals}
\begin{align}
\begin{split}
\boldsymbol{r}^{k} \coloneqq & \big[ \hat{p}_{1,0}^{k+1} - {p}_{1,0}^{k+1}, ..., \hat{p}_{H,T}^{k+1} - {p}_{H,T}^{k+1}, \hat{y}_{1,0}^{k+1} - {y}_{1,0}^{k+1}, ...,\\
& \hat{y}_{H,T}^{k+1} - {y}_{H,T}^{k+1}, \hat{p}_{1,0}^{k+1} - {p}_{1,0}^{k+1}, ..., \hat{q}_{H,T}^{k+1} - {q}_{H,T}^{k+1} \big] \\
=& \boldsymbol{r}_{p}^{k} + \boldsymbol{r}_{y}^{k} + \boldsymbol{r}_{q}^{k}, \label{eq:Residuals_primal}
\end{split} \\
\begin{split}
\boldsymbol{s}^{k} \coloneqq & \big[ \rho_{p} \left( {p}_{1,0}^{k+1} - {p}_{1,0}^{k} \right), ..., \rho_{p} \left( {p}_{H,T}^{k+1} - {p}_{H,T}^{k} \right), \\
& \rho_{y} \left( {y}_{1,0}^{k+1} - {y}_{1,0}^{k} \right) , ..., \rho_{y} \left( {y}_{H,T}^{k+1} - {y}_{H,T}^{k} \right), \\
& \rho_{q} \left( {q}_{1,0}^{k+1} - {q}_{1,0}^{k} \right), ..., \rho_{q} \left( {q}_{H,T}^{k+1} - {q}_{H,T}^{k} \right) \big] \\
=& \boldsymbol{s}_{p}^{k} + \boldsymbol{s}_{y}^{k} + \boldsymbol{s}_{q}^{k}, \label{eq:Residuals_dual}%
\end{split}
\end{align}
\end{subequations}
\normalsize
where (\ref{eq:Residuals_primal}) represent the violations of the coupling \cref{eq:Coupling_p,eq:Coupling_PVC,eq:Coupling_q} associated with the Augmented Lagrangian (\ref{eq:Lagrangian}) at the current iteration, and (\ref{eq:Residuals_dual}) represents the violation of the Karush-Kuhn-Tucker (KKT) stationarity constraints. 

The termination criteria are then given by:
%
\begin{equation} \label{eq:termination}
    {\lVert\boldsymbol{r}^{k}\rVert}_2 \leq \epsilon^{\text{pri}} \quad \text{and} \quad {\lVert\boldsymbol{s}^{k}\rVert}_2 \leq \epsilon^{\text{dual}},
\end{equation}
where $\epsilon^{\text{pri}}$ and $\epsilon^{\text{dual}}$ are feasibility tolerances determined by the following equations \cite{Boyd2011}:
%
\begin{subequations} \label{eq:Tolerances}
\begin{align}
\begin{split}
\epsilon^{\text{pri}} & = \sqrt{3HT} \epsilon^{\text{abs}} \\ 
& + \epsilon^{\text{rel}} \text{max} \big\{ {\lVert{\boldsymbol{\hat{p}}^{k}} + {\boldsymbol{\hat{y}}^{k}} + \boldsymbol{\hat{q}}^{k} \rVert}_2, {\lVert{\boldsymbol{p}^{k}} + {\boldsymbol{y}^{k}} + {\boldsymbol{q}^{k}}\rVert}_2 \big\}, \label{eq:Tolerances_primal}
\end{split}\\[10pt]
& \epsilon^{\text{dual}} = \sqrt{3HT} \epsilon^{\text{abs}} + \epsilon^{\text{rel}} {\lVert{\boldsymbol{\lambda}_{p}^{k}} + {\boldsymbol{\lambda}_{y}^{k}} + {\boldsymbol{\lambda}_{q}^{k}}\rVert}_2 \label{eq:Tolerances_dual}.%
\end{align}
\end{subequations}
\normalsize
%

%
%

In real-world applications, the standard ADMM may exhibit poor performance due to poor conditioning of the problem. Accelerated methods and adaptive penalty methods have been proposed to speed up the convergence of ADMM, but accelerated methods have been found to show no consistent or substantial improvement on non-convex AC OPF problems. Therefore, we adopt a residual balancing adaptive method \cite{Mhanna2019}, based on updating different $\rho$ independently, according to the relative magnitude of the primal and dual residuals \cite{Boyd2011}:

\begin{equation} \label{eq:ADMM_mu}
    \rho_{p}^{k+1} \coloneqq 
        \begin{dcases}
            \rho_{p}^{k} \left(1+\tau^{\text{incr}} \right) & \text{if} \quad {\lVert\boldsymbol{r}^{k+1}_{p}\rVert}_2 > \mu^{\text{incr}}{\lVert\boldsymbol{s}^{k+1}_{p}\rVert}_2,\\
            \rho_{p}^{k}{ \left (1+\tau^{\text{decr}} \right) }^{-1} \: & \text{if} \quad {\lVert\boldsymbol{s}^{k+1}_{p}\rVert}_2 > \mu^{\text{decr}}{\lVert\boldsymbol{r}^{k+1}_{p}\rVert}_2,\\
            \rho_{p}^{k} & \text{otherwise},
        \end{dcases} 
\end{equation}
\normalsize
where $\tau^{\text{incr}}, \tau^{\text{decr}} > 0$, $\mu^{\text{incr}}, \mu^{\text{decr}} > 1$ are parameters that tune how the differences between primal and dual residuals affect $\rho^{k+1}_{p}$. The same formula applies for $\rho_{y}$ and $\rho_{q}$, accordingly.


For our simulations, $\epsilon^{\text{abs}} = 0.001, \epsilon^{\text{rel}} = 0.01$, $\tau^{\text{incr}} = 1.15, \tau^{\text{decr}} = 0.9$, $\mu^{\text{incr}} = \frac{1}{1.15}, \mu^{\text{decr}} = 1.15$ for proportional and uniform dynamic redistribution, $\mu^{\text{incr}} = \frac{1}{0.8}, \mu^{\text{decr}} = 0.8$ for egalitarian redistribution.

Implementations without any form of adaptive methods have shown a very poor performance for any of the coordinated cases, where the number iterations can be orders of magnitude above the adaptive method. This is common among poorly conditioned non-convex problems \cite{Mhanna2019}. Furthermore, using separate weights $\rho_p, \rho_y$ and $\rho_q$ has also shown significant improvement over using a single weight, which is also a known characteristic of ADMM \cite{Boyd2011}.

The appropriate selection of the multi-objective weights $\alpha$ and $\beta$ may require close attention, and are proportional to the numerical conditioning of the problem. However, this is a general characteristic of multi-objective problems \cite{Boyd2011}. For $\beta$, values outside the optimal range may lead to an infeasible or slowly converging problem. The optimal range for $\beta$ was found to be from $1$ to $10$. 

Exceedingly high values of $\alpha$ can lead to slower convergence, and values too small fail to distribute the PVC across all prosumers. However, as noted in our results and mentioned in \cite{Liu_2020}, there is a trade-off between fairly distributing PVC and maximizing energy export, which can be seen as a cost of fairness. Small values for $\alpha$ can be used to obtain different fairness trade-offs. The optimal range of $\alpha$ was found to be from $50$ to $100$. Lower values of $\alpha$ fail to properly distributed the PVC across all prosumers, while higher values increase the number of iterations $k$.


\begin{figure}[!bp]
    \centering
    \resizebox{!}{3.25cm}{\begin{tikzpicture}

\filldraw [red] (0,0) circle (3pt);
\filldraw [black] (0.5,0.5) circle (3pt);
\filldraw [black] (1,1) circle (3pt);
\filldraw [black] (1.5,1.5) circle (3pt);
\filldraw [black] (2.3,1.5) circle (3pt);
\filldraw [black] (3.1,1.5) circle (3pt);
\draw (0.1,0.1) -- (0.5,0.5) -- (1,1) -- (1.5,1.5) -- (2.3,1.5) -- (3.1,1.5);

\filldraw [black] (1.2,0.5) circle (3pt);
\filldraw [black] (1.9,0.5) circle (3pt);
\filldraw [black] (2.4,1.0) circle (3pt);
\filldraw [black] (3.1,1.0) circle (3pt);
\filldraw [black] (3.8,1.0) circle (3pt);
\filldraw [black] (4.5,1.0) circle (3pt);
\draw (0.5,0.5) -- (1.2,0.5) -- (1.9,0.5) -- (2.4,1.0) -- (3.1,1.0) -- (3.8,1.0) -- (4.5,1.0);

\filldraw [black] (2.4,0.0) circle (3pt);
\filldraw [black] (3.1,0.0) circle (3pt);
\filldraw [black] (3.8,0.0) circle (3pt);
\filldraw [black] (4.3,-0.5) circle (3pt);
\filldraw [black] (4.8,-1.0) circle (3pt);
\filldraw [black] (4.8,-1.7) circle (3pt);
\draw (1.9,0.5) -- (2.4,0.0) -- (3.1,0.0) -- (3.8,0.0) -- (4.3,-0.5) -- (4.8,-1.0) -- (4.8,-1.7);

\filldraw [black] (2.4,-0.7) circle (3pt);
\filldraw [black] (2.4,-1.4) circle (3pt);
\filldraw [black] (1.7,-1.4) circle (3pt);
\filldraw [black] (1.0,-1.4) circle (3pt);
\draw (2.4,0.0) -- (2.4,-1.4) -- (1.7,-1.4) -- (1.0,-1.4);

\filldraw [black] (2.4,-2.1) circle (3pt);
\filldraw [black] (2.9,-2.6) circle (3pt);
\filldraw [black] (3.6,-2.6) circle (3pt);
\filldraw [black] (4.1,-3.1) circle (3pt);
\draw (2.4,-1.4) -- (2.4,-2.1) -- (2.9,-2.6) -- (3.6,-2.6) -- (4.1,-3.1);


\filldraw [black] (5.2,1.0) circle (3pt);
\filldraw [black] (5.7,0.5) circle (3pt);
\filldraw [black] (5.7,-0.2) circle (3pt);
\draw (4.5,1.0) -- (5.2,1) -- (5.7,0.5) -- (5.7,-0.2);

\filldraw [black] (5.9,1) circle (3pt);
\filldraw [black] (6.6,1) circle (3pt);
\filldraw [black] (7.1,1.5) circle (3pt);
\filldraw [black] (7.8,1.5) circle (3pt);
\filldraw [black] (8.3,2) circle (3pt);
\draw (5.2,1.0) -- (5.9,1) -- (6.6,1) -- (7.1,1.5) -- (7.8,1.5) -- (8.3,2);

\filldraw [black] (7.1,0.5) circle (3pt);
\filldraw [black] (7.8,0.5) circle (3pt);
\filldraw [black] (8.3,1) circle (3pt);
\filldraw [black] (8.8,1.5) circle (3pt);
\draw (6.6,1) -- (7.1,0.5) -- (7.8,0.5) -- (8.3,1) -- (8.8,1.5);

\filldraw [black] (8.3,0) circle (3pt);
\draw (7.8,0.5) -- (8.3,0);

\filldraw [black] (5.5,-2.3) circle (3pt);
\filldraw [black] (6,-2.8) circle (3pt);
\filldraw [black] (6.5,-3.3) circle (3pt);
\filldraw [black] (7,-2.8) circle (3pt);
\draw (4.8,-1.7) -- (5.5,-2.3) -- (6,-2.8) -- (6.5,-3.3) -- (7,-2.8);

\filldraw [black] (6,-1.7) circle (3pt);
\filldraw [black] (6.5,-1.2) circle (3pt);
\filldraw [black] (6.5,-0.5) circle (3pt);
\filldraw [black] (7.2,-0.5) circle (3pt);
\draw (5.5,-2.3) -- (6,-1.7) -- (6.5,-1.2) -- (6.5,-0.5) -- (7.2,-0.5);

\filldraw [black] (6.5,-2.3) circle (3pt);
\filldraw [black] (7,-1.7) circle (3pt);
\filldraw [black] (7.5,-1.2) circle (3pt);
\filldraw [black] (8.3,-1.2) circle (3pt);
\draw (6,-1.7) -- (6.5,-2.3) -- (7,-1.7) -- (7.5,-1.2) -- (8.3,-1.2);


\filldraw [black] (-0.75,-0.75) circle (3pt);
\filldraw [black] (-1.25,-1.25) circle (3pt);
\filldraw [black] (-1.75,-1.75) circle (3pt);
\filldraw [black] (-2.55,-1.75) circle (3pt);
\filldraw [black] (-1.75,-2.55) circle (3pt);
\draw (0, 0) -- (-0.75,-0.75) -- (-1.25,-1.25) -- (-1.75,-1.75) -- (-2.55,-1.75);


\filldraw [black] (-1.75,-0.75) circle (3pt);
\filldraw [black] (-2.55,-0.75) circle (3pt);
\filldraw [black] (-1.75,0.05) circle (3pt);
\filldraw [black] (-1.75,0.85) circle (3pt);
\filldraw [black] (-1.75,1.65) circle (3pt);
\draw (-1.25,-1.25) -- (-1.75,-0.75) -- (-1.75,0.05) -- (-1.75,0.85) -- (-1.75,1.65);

\filldraw [black] (-2.55,0.85) circle (3pt);
\filldraw [black] (-3.05,1.35) circle (3pt);
\filldraw [black] (-3.85,1.35) circle (3pt);
\filldraw [black] (-3.05,0.35) circle (3pt);
\draw (-1.75,0.85) -- (-2.55,0.85) -- (-3.05,1.35);

\filldraw [black] (-4.65,1.35) circle (3pt);
\filldraw [black] (-5.45,1.35) circle (3pt);
\filldraw [black] (-6.25,1.35) circle (3pt);
\filldraw [black] (-7.05,1.35) circle (3pt);
\draw (-3.05,1.35) -- (-4.65,1.35) -- (-5.45,1.35) -- (-6.25,1.35) -- (-7.05,1.35);

\filldraw [black] (-5.85,0.85) circle (3pt);
\filldraw [black] (-6.35,0.35) circle (3pt);
\filldraw [black] (-6.85,-0.25) circle (3pt);
\draw (-5.45,1.35) -- (-5.85,0.85) -- (-6.35,0.35) -- (-6.85,-0.25); 

\filldraw [black] (-3.85,0.35) circle (3pt);
\filldraw [black] (-4.65,0.35) circle (3pt);
\filldraw [black] (-5.15,-0.25) circle (3pt);
\filldraw [black] (-5.95,-0.25) circle (3pt);
\draw (-2.65,0.75) -- (-3.05,0.35) -- (-3.85,0.35) -- (-4.65,0.35) -- (-5.15,-0.25) -- (-5.85,-0.25);

\filldraw [black] (-3.35,-0.75) circle (3pt);
\filldraw [black] (-4.15,-0.75) circle (3pt);
\filldraw [black] (-4.65,-1.25) circle (3pt);
\filldraw [black] (-5.45,-1.25) circle (3pt);
\filldraw [black] (-6.25,-1.25) circle (3pt);
\draw (-1.75,-0.75) -- (-2.55,-0.75) -- (-3.35,-0.75) -- (-4.15,-0.75) -- (-4.65,-1.25) -- (-5.45,-1.25) -- (-6.25,-1.25);

\filldraw [black] (-3.35,-1.75) circle (3pt);
\filldraw [black] (-3.85,-2.25) circle (3pt);
\filldraw [black] (-4.65,-2.25) circle (3pt);
\filldraw [black] (-5.45,-2.25) circle (3pt);
\filldraw [black] (-6.25,-2.25) circle (3pt);
\filldraw [black] (-6.75,-1.75) circle (3pt);
\filldraw [black] (-6.75,-2.75) circle (3pt);
\draw (-2.55,-1.75) -- (-3.35,-1.75) -- (-3.85,-2.25) -- (-4.65,-2.25) -- (-5.45,-2.25) -- (-6.25,-2.25) -- (-6.75,-1.75);
\draw (-6.25,-2.25) -- (-6.75,-2.75);

\filldraw [black] (-5.15,-2.75) circle (3pt);
\filldraw [black] (-5.65,-3.25) circle (3pt);
\filldraw [black] (-6.45,-3.25) circle (3pt);
\draw (-4.65,-2.25) -- (-5.15,-2.75) -- (-5.65,-3.25) -- (-6.45,-3.25);

\filldraw [black] (-2.55,-2.55) circle (3pt);
\filldraw [black] (-1.75,-3.35) circle (3pt);
\filldraw [black] (-2.55,-3.35) circle (3pt);
\filldraw [black] (-3.35,-3.35) circle (3pt);
\filldraw [black] (-4.15,-3.35) circle (3pt);
\draw (-1.75,-1.75) -- (-1.75,-2.55) -- (-2.55,-2.55);
\draw (-1.75,-2.55) -- (-1.75,-3.35) -- (-2.55,-3.35) -- (-3.35,-3.35) -- (-4.15,-3.35);

\filldraw [black] (-2.25,2.25) circle (3pt);
\filldraw [black] (-3.05,2.25) circle (3pt);
\filldraw [black] (-3.85,2.25) circle (3pt);
\filldraw [black] (-4.65,2.25) circle (3pt);
\filldraw [black] (-5.45,2.25) circle (3pt);
\draw (-1.75,1.65) -- (-2.25,2.25) -- (-5.45,2.25);

\draw[red, dashed] (-7.55,2.75) -- (9.45,2.75) -- (9.45,-3.95) -- (-7.55,-3.95) -- (-7.55,2.75);
\draw[dashed] (-0.5,2.5) -- (9.3,2.5) -- (9.3,-3.8) -- (-0.5,-3.8) -- (-0.5,2.5);
\draw[blue, dashed] (-0.4,2.25) -- (4.95,2.25) -- (4.95,-3.65) -- (-0.4,-3.65) -- (-0.4,2.25);
\draw[magenta, dashed] (-0.3,2.075) -- (3.35,2.075) -- (3.35,-1.75) -- (-0.3,-1.75) -- (-0.3,2.075);
\draw[cyan, dashed] (-0.175,1.95) -- (2.1,1.95) -- (2.1,-0.85) -- (-0.175,-0.85) -- (-0.175,1.95);


\Large
\node[cyan] at (0,-0.6) {5};
\node[magenta] at (0,-1.45) {15};
\node[blue] at (0,-3.35) {25};
\node at (8.95,-3.5) {50};
\node[red] at (-7.10,-3.6) {100};

\end{tikzpicture}}
    \caption{\footnotesize 6-, 16-, 26-, 51- and 101-bus networks showing buses, lines, and the generator in red. The cyan, magenta, blue, black and red areas encompass, respectively, 5, 15, 25, 50 and 100 prosumers.}
    \label{fig:Network4}
\end{figure}
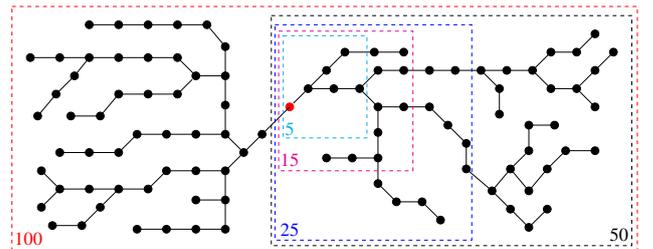

\subsection{Test networks} \label{Sec:Implementation-Network}
The proposed method was tested on low-voltage distribution feeders of different sizes (6, 11, 26, 51 and 101 buses) using two network topologies: tree topology with lateral spurs and line topology without lateral spurs.
The tree topology with lateral spurs is shown in Fig.~\ref{fig:Network4}. The network is split into subareas of different sizes with, respectively, 5, 10, 25, 50 and 100 prosumers shown as black dots (the red dot is the connection to the upstream medium voltage network modeled as an equivalent generator).
The line topology is simply a straight line feeder, as in  \cite{Peng2018}. Unlike in \cite{Peng2018}, we did not use the fat-tree topology because PV curtailment of a single prosumer does not affect other prosumers due to the connection to the medium voltage network in between.

A limit for the reactive power limit of $\SI{10}{\percent}$ was used in the simulations ($\overline{q}_h = 0.1 \overline{s}_h$). The inverter capacity $\overline{s}_h$ was set equal to the PV generation capacity, which required active power curtailment according to (\ref{eq:HEM_Inv_Cap}) when the inverter provided reactive power.
The voltage reference is $v_{r} = \SI{230}{\volt}$, and the voltage limits $\underline{v_{i}} = \SI{216}{\volt}$ and $ \overline{{v}_i} = \SI{253}{\volt}$ according to the Australian standard ($\SI{230}{\volt} \pm \SI{6}{\percent}/\SI{10}{\percent} $).
Voltages for VVC are $V_1 = \SI{216}{\volt}, V_2 = \SI{225}{\volt}, V_3 = \SI{244}{\volt}$ and $V_4 = \SI{253}{\volt}$. 

We used electricity demand and PV data with half-hourly resolution $\left( \Delta t = 0.5 \right)$ on a spring day (2011/11/07) from a recent Australian smart grid trial \cite{Ratnam2015}.
All prosumers have batteries with maximum capacity of $\SI{7.5}{\kilo\watt\hour}$, charging and discharging efficiency $\eta = 0.92$, and maximum charging or discharging rate of $\SI{3.75}{\kilo\watt}$. The time-of-use tariff ${c}^{\text{ToU}}_{t}$ was $\SI{0.12}{\$\per\kilo\watt\hour}$ off-peak (23:00 - 7:00), $\SI{0.22}{\$\per\kilo\watt\hour}$ shoulder (7:30 - 14:00, 20:00 - 22:30), and $\SI{0.52}{\$\per\kilo\watt\hour}$ peak (14:00 - 19:30), and the feed-in-tariff ${c}^{\text{FiT}}$ was $\SI{0.10}{\$\per\kilo\watt\hour}$. The battery state of charge ate the beginning of the horizon was set to $SoC_{h,0} = \SI{3}{\kilo\watt\hour}$. The APC of scenario B limits exporting up to $\SI{2}{\kilo\watt}$ for each prosumer.
The connection to the medium-voltage network is modeled as a generator with a quadratic cost function with ${c}_{2} =  \SI{0.025}{\$\per\kilo\watt\squared}$ for the three smaller networks and ${c}_{2} =  \SI{0.015}{\$\per\kilo\watt\squared}$ for the two bigger networks, and ${c}_{1} = {c}_{0} = 0$ for all networks.

\subsection{Test cases}
\label{sec:Simulation_scenarios}
We applied the three fair PVC methods on each network with three levels of PV penetration: low, medium, and high.
The low and medium cases do not require curtailment, so the choice of the PVC method is not important.
This gives 50 test cases in total, as summarized in Table~\ref{table:Topology}.
In summary, we considered the following seven scenarios: 
\begin{enumerate}[label=(\Alph*)]
    \itemsep0em 
    \item \textit{Uncoordinated}: prosumers act in a selfish manner, seeking the best cost according to connection tariffs.
    \item \textit{Uncoordinated, VVC, fixed APC}: inverters have VVC enabled. The active power export limit is set to $\SI{2}{\kilo\watt}$ for each prosumer.
    \item \textit{Coordinated}: the standard DER coordination problem \cite{Gebbran_AAMAS,Attarha2020} is utilized.
    \item \textit{Coordinated, VVC}: \eqref{eq:Central_OPF_2nd} implemented in a distributed manner.
    \item \textit{Coordinated, VVC, PVC \#1}: implementation of \eqref{eq:ADMM} using an egalitarian PVC redistribution.
    \item \textit{Coordinated, VVC, PVC \#2}: implementation of \eqref{eq:ADMM}, using a proportional PVC redistribution.
    \item \textit{Coordinated, VVC, PVC \#3}: implementation of \eqref{eq:ADMM}, using an uniform dynamic PVC redistribution.
\end{enumerate}

 
\section{Simulation results} \label{Sec:Results}
We present the results in two parts: we first discuss in detail the results for  the 51-bus network  with  tree topology and 50 prosumers (Fig.~\ref{fig:Network4} black), followed by a comparison across all 50 cases focusing on the algorithmic performance only.

\subsection{50-prosumer network with tree topology} \label{Sec:Results_AB}
\begin{figure*} 
    \centering
  {\includegraphics[width=1\linewidth]{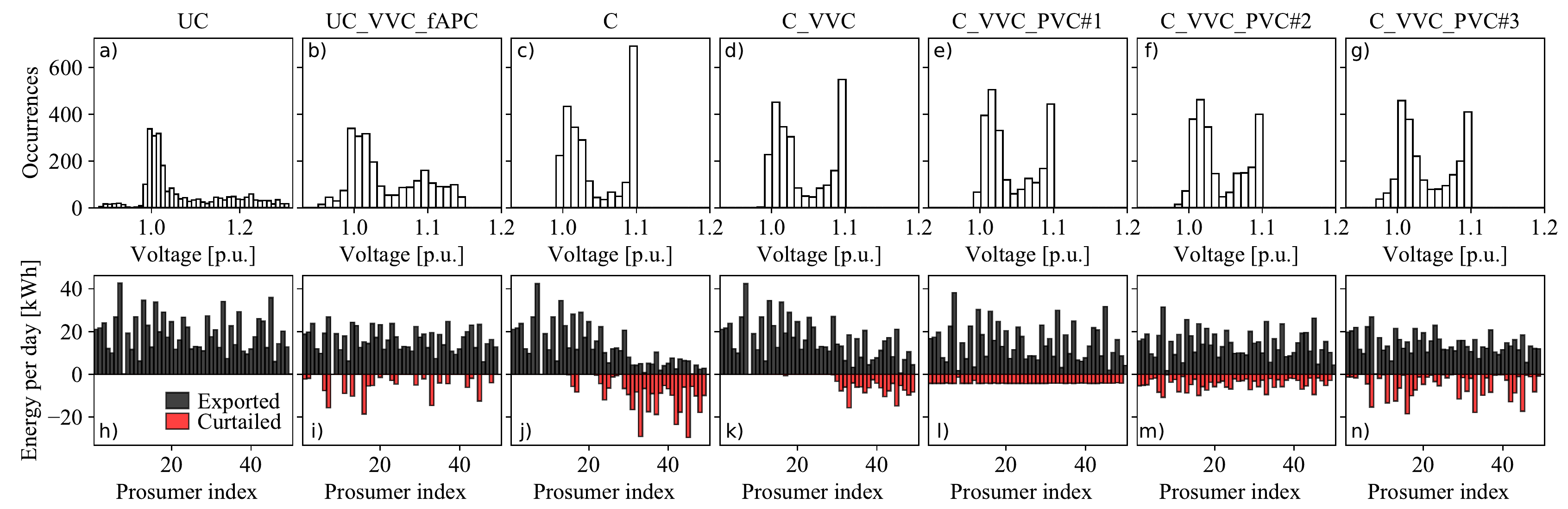}} 
  \caption{Results for the 50-prosumer network with tree topology under the seven proposed scenarios. Figures a-g show the distribution of voltages $v_{h,t}$ over a day, and figures h-n show the amount of energy exported $\sum_{t} \boldsymbol{p}_h^-$, in black, and curtailed $\sum_{t} \boldsymbol{y}_h$, in red, for each prosumer over a day.}
  \label{fig:Results_42} 
\end{figure*}

Fig.~\ref{fig:Results_42} shows a comparison of all seven scenarios detailed in Section \ref{sec:Simulation_scenarios}, focusing on the occurrence of voltage violations (top graphs) and the curtailed PV generation (bottom). 

\textbf{Scenarios A and B}: As expected in networks with a high PV penetration, the voltage stress is very high. When uncoordinated, prosumers exceed the voltage limits: over $1.3$ p.u., and under $0.93$ p.u. as seen in Fig.~\ref{fig:Results_42}a. 
In scenario B, shown in Fig.~\ref{fig:Results_42}b, we still have over-voltage because there is no coordination between agents. Due to the APC, no prosumer exceeds $\SI{25}{\kilo\watt\hour}$ in a day of exported energy, as seen in Fig.~\ref{fig:Results_42}l.

\textbf{Scenarios C and D}: When coordinating prosumers, and curtailing PV generation, we can ensure proper operation within network limits. This is shown in  Fig.~\ref{fig:Results_42}c, where voltages lie within the required limits $0.96$ - $1.1$ p.u. Note, however, that the PVC depends on the electrical distance; that is, prosumers at the end of the feeder are curtailed more. Fig.~\ref{fig:Results_42}j shows that over half of the prosumers are curtailed for the most part of the day.
If we enable the VVC for all inverters, we have less voltage stress on the network, as shown in Fig.~\ref{fig:Results_42}d, but we still have an unfair (albeit smaller) PVC, shown in  Fig.~\ref{fig:Results_42}k.

\textbf{Scenario E}: The curtailment is spread across all prosumers roughly equally (Fig.~\ref{fig:Results_42}l). There are minimal fluctuations of prosumers curtailment, since some prosumers may not export at any given time interval $t$. As proposed, prosumers who are not exporting energy at $t$ are not curtailed. 
Observe in Fig.~\ref{fig:Results_42}e how the voltage stress in the network is reduced.

\textbf{Scenario F}: The proportional redistribution spreads the PVC across all prosumers in proportion to the exported power at each interval $t$ (Fig.~\ref{fig:Results_42}m). Once again, there are small variations for prosumers curtailment, because the contributions of individuals are different at each $t$.  
Nevertheless, the proportion of energy curtailed versus the sum of curtailed and exported energy ranges between 15\% and 30\%, which can be attributed to different rooftop PV orientations and demand profiles. 

\textbf{Scenario G}: The uniform dynamic redistribution curtails more aggressively prosumers who export more. Prosumers who export less are curtailed very little or not at all, as shown in Fig.~\ref{fig:Results_42}n. The reduction in voltage stress on the network (Fig.~\ref{fig:Results_42}g) is comparable to Scenarios E and F.

\subsubsection{General observations} \label{Sec:Results_GeneralCons}

Observe in Table~\ref{table:Tolerances} that the total amount of curtailed energy $\sum y_{h,t}$ is higher among cases with a fair PVC (E, F, and G) when compared to a similar case without a fair PVC (D). The fairness penalty drives the pattern of curtailment away from the most efficient configuration of Scenario D. However, the results show a smaller total curtailed energy when compared to the coordinated case without VVC (C). This is, in turn, proportional to the maximum reactive power $\overline{q}_h$.\footnote{Our selected value $\overline{q}_h = 0.1 \overline{s}_h$ is below all Australian DNSP's requirements (which range from $0.3$ to $0.6$) \cite{CleanEnergyCouncil}, and is already enough to demonstrate the impact of VVC in networks with a high DER penetration.}

The impact of fair curtailment and the reactive power compensation using VVC in the original cost function \eqref{eq:Central_OPF_1st} is shown in Table~\ref{table:Tolerances} (columns $F_{\%C}$ and $F_{\%D}$). $F_{\%C}$ shows the difference of $F(\boldsymbol{x},\boldsymbol{z})$ between any given scenario and the coordination without neither VVC nor fair PVC (C), while $F_{\%D}$ shows the same difference with respect to Scenario D (with VVC but without fair PVC). The worst case scenario is Scenario G, where the increase in the cost function is $\SI{15.3}{\percent}$ compared to the optimal result of the DOPF without considering fair PVC (D). As discussed previously, all scenarios outperform the original case without VVC (C).

The difference of total curtailed energy $\sum y_{h,t}$ between scenarios E and F is negligible, but Scenario G has a larger amount of curtailment. This shows that the uniform dynamic PVC redistribution reduces the total exported energy the most. 
However, the uniform dynamic PVC redistribution affects 
the lowest-exporting prosumers the least. Conversely, the egalitarian redistribution spreads the PVC roughly equally among all energy-exporting prosumers, curtailing lowest-exporting prosumers the most. The proportional redistribution strikes a middle ground between these two.

The last column of Table~\ref{table:Tolerances} shows the average coefficient of variation of the curtailment $\overline{\mathrm{CV}} y_{h,t}$ during the daylight hours of a day. 
The variation in scenarios C, D and G are very high, which indicates the curtailment is spread unevenly. This is because prosumers are curtailed according to electrical distances for scenarios C and D, and higher exporting prosumers are curtailed more aggressively in scenario G. Scenario E shows a very low variation (evenly spread curtailment), while Scenario F lies somewhere between these two extremes.

\begin{table}[t]
\small
\renewcommand{\arraystretch}{1.1}
\centering
\caption{Number of iterations, parallel computation time, sum of PVC, deviation from the optimal cost of the solution without VVC (scenario C) and with VVC (scenario D), and average coefficient of variation of PVC, for all distributed scenarios (C-G).}
\label{table:Tolerances}
\begin{tabular}{|c|c|c|c|c|c|c|}
\hline
Scen. & $k$ & $\text{t}_\text{Par} [\SI{}{\second}]$ & $\sum y_{h,t}$  [\SI{}{\kilo\watt\hour}] & $F_{\%C}$ & $F_{\%D}$ & $\overline{\mathrm{CV}} y_{h,t}$ \\
\hline
C & 27 & 157.8 & 311.9 & N/A & 23.8 & 1.48 \\ 
D & 25 & 149.0 & 145.2 & -31.2 & N/A & 1.87 \\ 
E & 19 & 156.4 & 201.3 & -15.4 & 12.0 & 0.19 \\ 
F & 24 & 171.3 & 194.9 & -18.5 & 9.7 & 0.92 \\ 
G & 19 & 165.6 & 227.0 & -11.1 & 15.3 & 1.85 \\ 
\hline
\end{tabular}
\end{table}

\subsubsection{Computational performance} \label{Sec:Results_Comp}
The four distributed scenarios were also simulated in their centralized form (\ref{eq:Central_OPF_PVC}). Bonmin \cite{BONMIN} is used to solve Problem~\eqref{eq:Central_OPF_PVC}. 
%
However, these simulations became intractable. The simulation did not converge for any of the VVC scenarios for the first part of the results (D, E, F, and G, 50-prosumer network) in more than 72 hours. The number of binary variables resulting from the VVC constraint was over a thousand for this network. 


The number of iterations $k$ for the algorithm to converge 
was actually smaller for scenarios E, F and G, which is shown in Table~\ref{table:Tolerances}. The total parallel computation time $\text{t}_\text{Par}$ (the sum of the slowest parts when solving (\ref{eq:ADMM})) is slightly larger, resulting from a longer computation time per iteration due to the larger number of variables. The number of iterations is comparable to other approaches that use prosumer-based decomposition \cite{Scott2019,Andrianesis2019,Attarha2020}.


\subsubsection{Network and prosumer power flows} \label{Sec:Results_PF_Figs}



\begin{figure} 
    \centering
  {\includegraphics[width=1\linewidth]{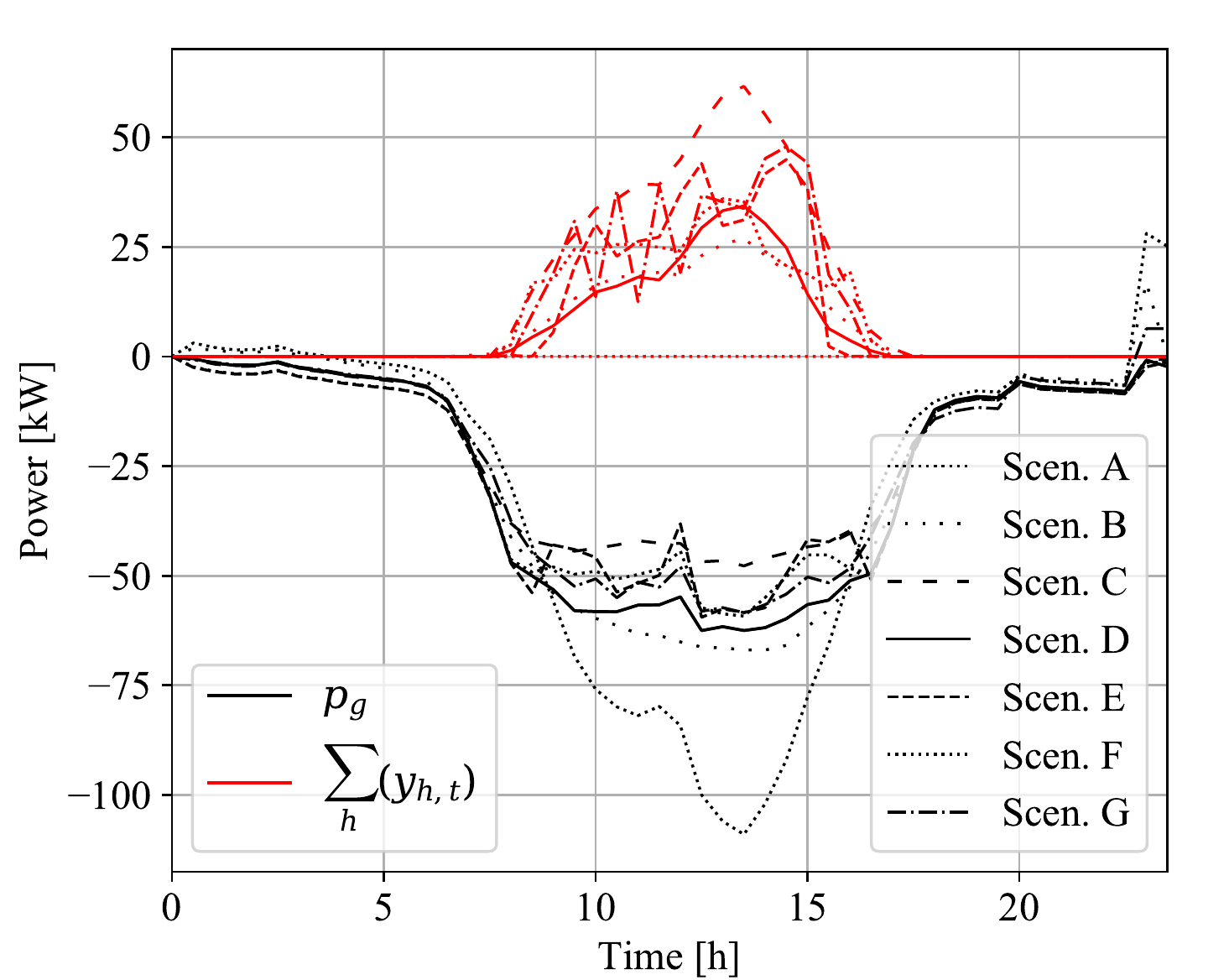}}
  \caption{Cumulative PV curtailments $y_{h,t}$ (red) and cumulative power flows through the distribution transformer $p_{g,t}$ (black) for the 50-prosumer network with tree topology for all scenarios.}
  \label{fig:Results_Power_Flow} 
\end{figure}


Fig.~\ref{fig:Results_Power_Flow} shows cumulative PV curtailments $y_{h,t}$ and cumulative power flows through the distribution transformer $p_{g,t}$.
Observe that the power flow through the distribution transformer varies little across scenarios, with slightly more PV curtailment in cases with PVC redistribution, as discussed in Section~\ref{Sec:Results_GeneralCons}.

\begin{figure*}[]
    \centering
  {\includegraphics[width=1\linewidth]{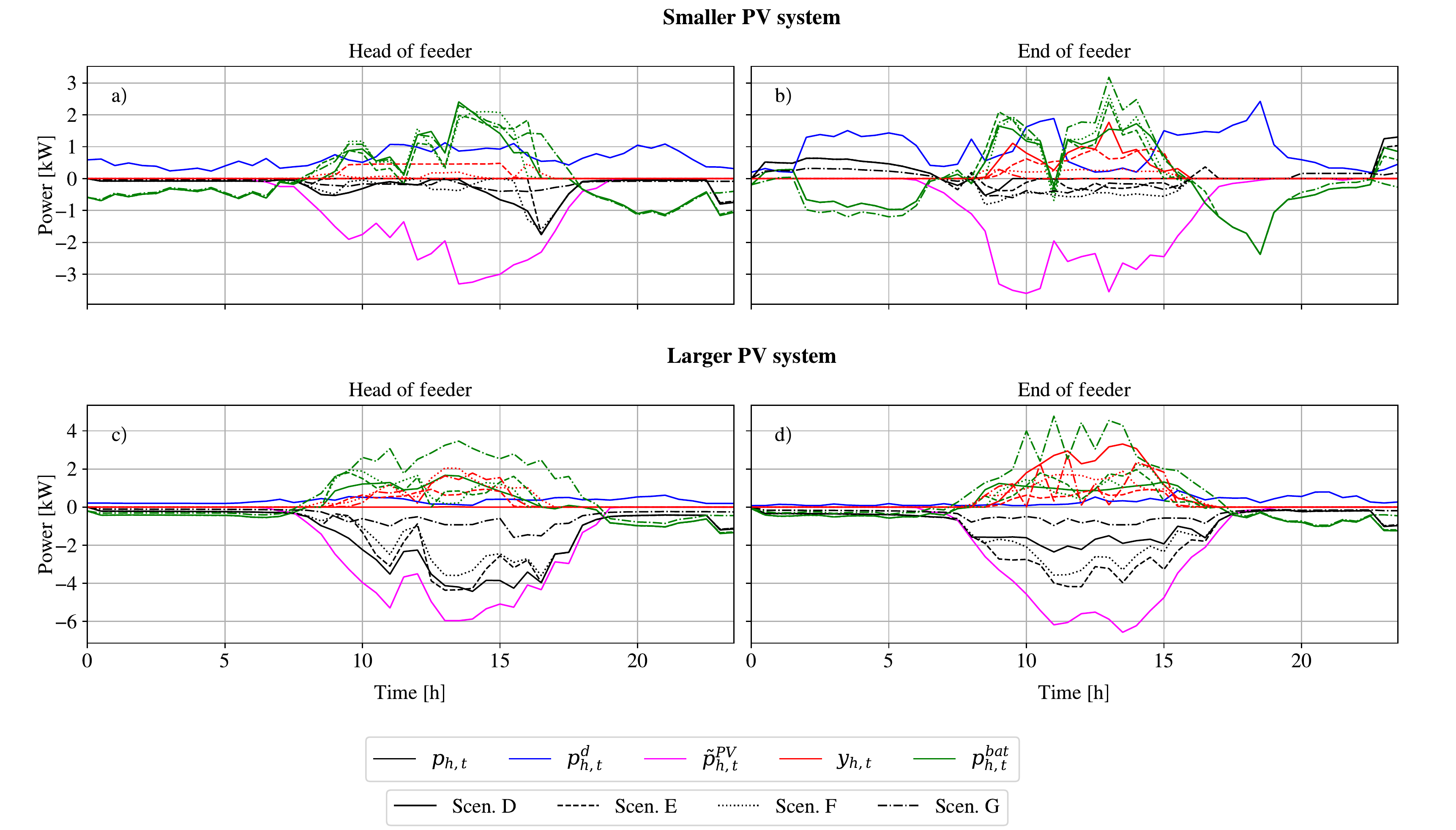}}
  \caption{Power flows for four selected prosumers for the 50-prosumer network with tree topology for scenarios D, E, F and G. Prosumers are selected based on the size of their PV system (small in cases a and b, big in cases c and d) and their location on the feeder (close to the feeder head in cases a and c, at the end of the feeder in cases b and d).}
  \label{fig:Prosumers_Power_Flow} 
\end{figure*}

Fig.~\ref{fig:Prosumers_Power_Flow} shows power flows for scenarios D, E, F and G for four selected prosumers: two at the feeder head and two at the end of the feeder, and two with a bigger and two with a smaller PV system. 
Prosumers close to the feeder head (Figs.~\ref{fig:Prosumers_Power_Flow}a and \ref{fig:Prosumers_Power_Flow}c) show no PVC in scenario D; that is, they are only curtailed when a fair PVC scheme is in effect. 
Meanwhile, prosumers located at the end of the feeder (Figs.~\ref{fig:Prosumers_Power_Flow}b and \ref{fig:Prosumers_Power_Flow}d) are curtailed more in scenario D (no PVC redistribution) when compared to scenarios E, F and G.
Prosumers with bigger PV systems (Fig.~\ref{fig:Prosumers_Power_Flow}c and \ref{fig:Prosumers_Power_Flow}d) are curtailed more in scenarios F (proportional) and G (uniform dynamic) when compared to scenario E (egalitarian). Conversely, prosumers with smaller PV systems (Figs.~\ref{fig:Prosumers_Power_Flow}a and \ref{fig:Prosumers_Power_Flow}b) are curtailed more in scenario E than in scenario F, and are not curtailed at all in scenario G.
These results agree with the results in \cite{Liu_2020}, both in terms of the location of the prosumers and the size of their respective PV systems. 

\begin{table}[!ht]
\renewcommand{\arraystretch}{1.0}
\small
\centering
\caption{Number of iterations for fifty simulations with different PV penetration, curtailment redistribution method, network size and topology.}
\label{table:Topology}
\begin{tabular}{|c||c|c|c|}
\hline
\thead{PV Penetration \\ \& Method}& \thead{Network \\ Size} & \thead {$k$ \\ (Line topology) \vspace{10pt}} & \thead{ $k$ \\ (Tree topology with\\radial spurs) } \\
\hline
\multirow{5}{*}{Low} & 5 & 13 & 15 \\
& 15 & 12 & 13 \\
& 25 & 26 & 17 \\
& 50 & 27 & 16 \\
& 100 & 26 & 18 \\
\hline
\multirow{5}{*}{Medium} & 5 & 23 & 8 \\
& 15 & 7 & 6 \\
& 25 & 29 & 7 \\
& 50 & 30 & 8 \\
& 100 & 28 & 12 \\
\hline
\multirow{5}{*}{\makecell{High --\\ Egalitarian}}& 5 & 28 & 32 \\
& 15 & 31 & 22 \\
& 25 & 35 & 19 \\
& 50 & 36 & 19 \\
& 100 & 45 & 25 \\
\hline
\multirow{5}{*}{\makecell{High --\\ Proportional}} & 5 & 22 & 18 \\
& 15 & 33 & 18 \\
& 25 & 29 & 16 \\
& 50 & 27 & 24 \\
& 100 & 27 & 19 \\
\hline
\multirow{5}{*}{\makecell{High --\\ Uniform \\ Dynamic}} & 5 & 17 & 12 \\
& 15 & 16 & 15 \\
& 25 & 20 & 20 \\
& 50 & 23 & 19 \\
& 100 & 22 & 22 \\
\hline

\end{tabular}
\end{table}
\subsection{Comparison of algorithmic performance of all cases}
The number of iterations $k$ for each of the 50 test cases is shown in Table~\ref{table:Topology}.
Observe that the the number of iterations is not directly related to the size or the type of the network, which is also shown in \cite{Gebbran_SGES}.
In general, the line topology required a slightly larger number of iterations when compared to the tree topology with radial spurs, which agrees with the results in \cite{Peng2018}. Note, however, that \cite{Peng2018} uses a component-based decomposition so the total number of iterations is significantly higher. By contrast, we solve the network subproblem in one piece, which reduces the number of iterations significantly at the expense of the increase in the computational burden in each iteration.

The primal residuals ($\boldsymbol{r}_p, \boldsymbol{r}_y, \boldsymbol{r}_q$) across all fifty simulations were on average less than $\SI{0.0015}{\kilo\watt}$, and their maximum values were $\SI{0.050}{\kilo\watt}$, $\SI{0.048}{\kilo\watt}$ and $\SI{0.068}{\kilo\watt}$ for $\boldsymbol{r}_p, \boldsymbol{r}_y, \boldsymbol{r}_q$, respectively. These values are within the same range as in \cite{Scott2019}, which has been successfully deployed in a real-life trial.
Note that a rolling-horizon implementation used in \cite{Scott2019} would likely result in a faster convergence due to the warm start of each subsequent horizon.



\subsection{{\color{black}Practical implementation considerations}} \label{Sec:Deployment}

\subsubsection{{\color{black}Convergence}}{\color{black}It is known that distributed optimization approaches for the non-convex AC OPF problem do not theoretically guarantee a globally optimal solution \cite{Molzahn2017}. However, as discussed in \cite{Molzahn2017} and demonstrated extensively in \cite{Mhanna2019,Mhanna2017ACD}, ADMM has been numerically shown to converge on a wide range of OPF cases. Our simulations featuring ten different network configurations under three different levels of PV penetration, for a total of fifty different simulation scenarios, all exhibit good convergence, as evidenced by the vanishing primal and dual residuals. 
In other words, the KKT conditions are satisfied since the primal and dual residuals converge within the established tolerance \cite{Boyd2011}. It is noteworthy that the KKT conditions are \emph{necessary} to ascertain a (locally) optimal point in a non-convex problem, and \emph{sufficient} to ascertain a globally optimal point in a convex problem. The robustness shown by the results of our ADMM-based algorithm for different test cases agree with previous results using ADMM for non-convex DOPF problems, whether in large systems decomposed by areas \cite{Guo_2017}, in element-wise decomposition \cite{Mhanna2017ACD} or in prosumer-based decomposition \cite{Scott2019, Andrianesis2019,Attarha2020,Gebbran_SGES}.}

\subsubsection{{\color{black}Parameter Settings}}{\color{black}Furthermore, parameters need to be reset between different network sizes/topologies and curtailment redistribution methods (beyond the discussion in subsection 6.1). This is a known requirement for non-convex distributed optimization problems in general, in particular for ADMM \cite{Boyd2011, Mhanna2017ACD, Mhanna2019}. However, once the parameter settings are established for a given system and redistribution method, any scenario can be simulated under the same algorithmic parameter settings. This means that for implementation purposes on a single network, the distribution system operator would only need to tune the parameters once for each desired curtailment redistribution method (or only once, if using a single curtailment method), and proceed to operating the algorithm with that set of parameters for all scenarios.}

\subsubsection{{\color{black}Hardware implementation}}{\color{black}It has been shown that edge computing devices with low processing power are able to compute the prosumer subproblem quickly (in less than a second), which does not impede the deployment of this algorithm in actual distributed settings \cite{Gebbran_SGES}. Alternatively, different distributed computation archetypes may be employed, such as grouping a small number of prosumers within an area and assigning one local computing agent to handle their subproblems, as has been implemented in a real-world trial in Australia \cite{Scott2019}. }

\subsubsection{{\color{black}Communication Requirements}} {\color{black}The proposed algorithm requires communication between the aggregator and the prosumers. 
The size of the messages exchanged between the agents is smaller than 2 KB, which can be accommodated using the existing last-mile network technologies, such as 4G, or networks tailored for the Internet of Things \cite{Gebbran_AUPEC} such as LTE-M, NB-IoT and EC-GSM-IoT.
These technologies offer sufficiently small latencies relative to the computation time of the DOPF network subproblem so that the communication delay accounts for less than 5\% of the total computation time \cite{Gebbran_SGES}. 
Economical aspects and limiting factors such as low area coverage and poor internet connection would probably play a bigger tole in choosing the appropriate network technology.}

%
%
%

\section{Conclusion} \label{Sec:Conclusion}
This paper has proposed a novel approach for {\color{black}distributed energy resources} (DER) coordination via an AC {\color{black}distributed optimal power flow} inspired by {\color{black}the alternating direction method of multipliers} (ADMM). Compared to existing methods, the proposed solution includes a standard inverter {\color{black}Volt-Var Control} (VVC) function for reactive power compensation and a fair {\color{black}PV curtailment} (PVC) distribution among prosumers, which have previously been only addressed in isolation. The resulting problem is a {\color{black}mixed-integer nonlinear problem} due to the piecewise linear VVC function and the non-convex network problem, and is shown to be intractable if solved centrally. The proposed distributed approach, on the other hand, is tractable because it avoids using integer variables. Moreover, including fair PVC and inverter VVC in the DER coordination problem retains its privacy-preserving characteristic and the computational efficiency, both in terms of the computation time and the number of iterations. 

We have demonstrated the effectiveness of the proposed method on 50 different test cases using low-voltage distribution networks of different sizes and topologies, and with different PV penetration levels. The simulation results on a large number of diverse cases demonstrate the robustness of the proposed approach, even though ADMM does not offer convergence guarantees for non-convex problems.

We have proposed three different approaches for PVC redistribution, each adopting different notions of fairness. The results demonstrate that the combination of inverter VVC and PVC effectively reduces voltage stress on the network without penalizing prosumers at the end of the feeder. This stands in contrast to conventional DER coordination approaches that curtail electrically-distant prosumers the most. The overall increase in the cost compared to an optimal solution with unfair allocation PVC (with VVC) is $\SI{15.3}{\percent}$ in the worst-case scenario.

The proposed method requires frequent message exchange between prosumers and the central coordinator, which requires a communication network with sufficiently low latency and high enough bandwidth. {\color{black} Current network technologies have a sufficiently small latency, but} the impact on the communication network traffic {\color{black} for a very large number of agents needs further research}.






\bibliographystyle{model3-num-names}

\bibliography{cas-refs}





\end{document}